# Indium-Tin-Oxide for High-performance Electro-optic Modulation

Zhizhen Ma[1], Zhuoran Li[1], Behrouz Movahhed Nouri[1,3], Ke Liu[1,2], Chenran Ye[1], Hamed Dalir[1,3], Volker J. Sorger[1*]

[1]Department of Electrical and Computer Engineering, School of Engineering and Applied Science, George Washington University, Washington, DC 20052, USA
[2]The Key Laboratory of Optoelectronics Technology, Ministry of Education, Beijing University of Technology, Beijing 100124, P.R. China
[3]Optelligence LLC, 10703 Marlboro Pike, Upper Marlboro, MD, 20772, USA

[*]Corresponding author: sorger@gwu.edu

**Abstract:** Advances in opto-electronics are often led by discovery and development of materials featuring unique properties. Recently the material class of transparent conductive oxides (TCO) has attracted attention for active photonic devices on-chip. In particular Indium Tin Oxide (ITO) is found to have refractive index changes on the order of unity. This property makes it possible to achieve electro-optic modulation of sub-wavelength device scales, when thin ITO films are interfaced with optical light confinement techniques such as found in plasmonics; optical modes are compressed to nanometer scale to create strong light-matter-interactions. Here we review efforts towards utilizing this novel material for high-performance and ultra-compact modulation. While high performance metrics are achieved experimentally, there are open questions pertaining the permittivity modulation mechanism of ITO. Furthermore, we show that a footprint-saving waveguide inline cavity can enhance obtainable extinction-ratio to insertion-loss ratios by about one order of magnitude over non-cavity based version. Moreover, we offer a speed analysis that shows that the device is resistance limited, but not capacitance or drift-carrier limited. Interestingly, two bias options exist for ITO and we find that a side-connection enables devices that should in principle enable several hundred of GHz fast devices, using our routinely achievable ITO film resistivities. Finally, we offer a brief discuss about footprint savings of compact ITO modulators showing a 3-orders of magnitude smaller footprint over Silicon photonic MZI-based modulators. Such compact performance can play a key role in emerging ASICs such as photonic tensor core accelerators for machine learning. Lastly, we review a variety of optical and electrical properties of ITO for different processing conditions, and show that ITO-based plasmonic electro-optic modulators have the potential to significantly outperform diffraction-limited devices.

## 1. Introduction

A potential viable way of fulfilling both size and power requirement for future photonic integrated circuit (PIC) technology lies in down-scaling opto-electronic devices beyond the diffraction limit of light [1-4]. The advantage of such sub-diffraction limited photonics is two-fold; reduced optical power requirements and physical device size. To elaborate on this, while being physically compact, the optical mode confinement of such components can strongly enhance light matter interactions (LMI) [5,6], which in terms can reduce required drive power to obtain the desired effect, e.g. signal modulation, optical non-linearities [7,8]. In order to address these demands, photonic components and even circuits based on surface plasmon polaritons (SPPs), collective oscillations of electrons at metal-dielectric interfaces are thought of a solution for nanoscale PICs [9]. However, while SPP-based schemes have been explored before, many do not offer both sub-wavelength confinement beyond the diffraction limit and long enough interaction lengths making these designs often unsuitable towards nanoscale photonic integration [10, 11]. As a result, the use of plasmonics for photonic on-chip solutions, in particular for optical interconnects remained uncertain until recently. However, emerging materials such as the recently explored transparent conductive oxides (TCOs) show promise to deliver greatly enhanced functionality and performance.

One possible path for future photonic technology is to decouple passive light routing from active light-manipulation leading to hybrid integration strategies [8]. To this end, a low-loss platform would build the backend data link, e.g. Silicon-on-insulator (SOI) based waveguides [12], while SPP-based components offer compact and efficient light manipulation [5, 6, 13-16]. Therefore, these components are required to fulfil two design criteria; strong mode confinement and a sufficiently long propagation length to utilize the mode towards light manipulating effect or signal [17, 18]. Both criteria combined can be used to form a figure-of-merit (FOM) suitable for comparing design concepts towards highly scaled integrated nanophotonic building blocks as will discuss below. A remaining challenge, however, is then given by the need to reduce the optical coupling loss arising from the optical mode size mismatch between the diffraction limited and sub-diffraction limited modes in the passive and active regions of the PIC, respectively [19]. In addition to the LMI enhancements, the choice of the optical active material is critical; towards enhancing the device performance for on-chip photonic building blocks, a variety of novel materials are emerging in recent years. These materials inhibit unique properties for nanoscale opto-electronics such as high carrier density for a tuneable permittivity. Here, we discuss a particular material Indium-Tin-Oxide (ITO), which belongs to the family class of TCOs. By reviewing material fundamentals and recent work in nanophotonics, we show that using ITO and nanophotonic field enhancement techniques produce the aforementioned high FOM devices as exemplified by the first λ-scale electro-optic modulators (EOM) [20].

## 2. Transparent Conductive Oxides Used in Photonics

TCOs are widely used for their optically transparent yet electrically conductive properties. As such, TCO's found increasing interest in optoelectronic technologies such as for displays (i.e. flat panel displays) and energy conversion (i.e. photovoltaics, architectural and window glass). The discovery of TCOs dates back about one century, where Badeker found the resistivity of CdO to be as low as $1.2\times10^{-3}$ Ω·cm. Note, this is about one order of magnitude higher than that of ITO films – the widely used TCO material today [21]. Tin dioxide ($SnO_2$) made technological progress when scientists applied it as a transparent heating layer in the airplane cockpit windows [22]. The broad industrial utilization of TCO materials occurred at the end of the 1960s, when infrared light filters composed of Tin or $In_2O_3$ were used on low-pressure sodium discharge lamps for increasing the lamp efficiency by reducing heat losses [23, 24]. Afterward with the advent of flat-panel display technology around 1970, ITO became the most commonly used TCO material for transparent electrodes [25]. Up to date, ITO is the TCO material having with the lowest resistivity (i.e. order of $1\text{-}2\times10^{-4}$ Ω·cm), which is widely utilized as the transparent electrodes in flat-panel displays [28, 29].

Table 1 summarizes the most promising mechanisms of modulation and its corresponding materials. In 2010, the Atwater group found the characteristic of unity order refractive index change for TCO materials, and gave ways to utilize TCO films as an active material in electro-optic modulators, which could lead to a new generation of fast, on-chip, nanoscale devices with unique capabilities. The mechanism for TCO films as an active material in EOMs resembles the electrostatics found in metal-oxide-semiconductor (MOS) capacitors, a bias voltage can change the carrier concentration in an accumulation (or inversion) layer of a modulator device. This carrier change has been found to result in a refractive index change, therefore an intensity modulation could be achieved. This modulation mechanism of carrier concentration has been deployed in semiconductor modulators for many years. As an example, carrier-effect based Silicon optical modulators (e.g. carrier-depletion) may reach a modulation speed of over 50 Gbps [30]. However, these devices still suffer a weak light-matter-interaction (i.e. refractive index change of $2 \times 10^{-3}$ at a wavelength of 1.55 μm with $10^{18}$ carriers/cm$^3$ [31]) and then a larger device footprint. The behavior of phase transition (i.e. a semiconductor to metal phase transition) is observed in some smart materials, presenting a noticeable permittivity change between the two phases. For example, vanadium dioxide ($VO_2$) has been studied extensively in the last decade because of its large, reversible change in its electrical, optical, and magnetic properties at a temperature close to room temperature. At room temperature $VO_2$ behaves as a semiconductor with a band gap of ~1 eV. At temperatures higher than 68°C $VO_2$ passes through an abrupt transition to a metallic state. This process is reversible when lowering the temperature below 65°C ($VO_2$ becomes semiconductor again). Taking advantage of the phenomenon, a compact $VO_2$-based absorption modulator at 1550 nm wavelength on a SOI platform can be realized through the stimuli of substrate heating [32]. The thermo-optic effect results from a temperature dependent refractive index of a material, such as in tuning the refractive index of Silicon via resistive heating (i.e. ~$1.8\times10^{-4}$/K of thermo-optic coefficient of Silicon) [33]. Another widely used thermo-optic effect material is polymer, which has the advantages of low density and refractive index, less dispersion, cheap manufacturing cost, easy to make complex shapes, and good electrical insulation. However, this type of materials suffers shortcomings such as limited temperature usability, degraded time dependence of properties (e.g. electro-optic polymers), and fatigue and solvent sensitivity. Although polymer-based thermo-optic modulators (and switches) have several drawbacks such as significant power consumption, slow transition time (usually > 1 μs), the mechanism shows operational devices and sub-systems on-chip [34, 35]. Phase transition and thermal-optic effect based modulators can provide significant response, however the time frame is high and allows for an operation speed on the order of MHz [36].

TCOs such as ITO, GZO (ZnO: Al), and AZO (ZnO: Ga) have very similar properties and allow for an efficient change of carrier concentration. As an outstanding representative of TCO materials used for EOMs, ITO exhibits its unique advantages, such as epsilon-near-zero (ENZ) in the near infrared regime and electrically tunable permittivity [37, 38]. A recently explored example of MOS-based EOM is an ITO-$SiO_2$-Au stack in a Silicon-on-Insulator waveguide [20]. The effective index of ITO changes from being a dielectric to a quasi-metallic state when a voltage bias is applied. Such a modulator exhibits a low insertion loss of ~1 dB for the 5 μm long device, a sub-wavelength compact size of 0.78 λ, and the broadband operation of > 500 nm with a non-resonant MOS mode [39]. Though TCO materials have a strong electro-optical response compared to the aforementioned materials (Table 1). A large index change (real part) is also accompanied with a significant extinction coefficient change (imaginary part of index) making TCOs often rather lossy due to the Kramers-Kronig relations. However, since the carrier concentration of TCOs may be selectively controlled in a post-processing, local resistivity tuning step. This allows separating the optically active material from the passive electronic, and may be realized by a plasma treatment process to optimize the resistivity of TCO layer as discussed in detail below. In the remaining sections we specifically focus on material, processing, and electro-optic modulation details for the material ITO.

## 3. ITO Material Science and Processing Fundamentals

ITO sputtering targets typically consist of 10% $SnO_2$ to 90% $In_2O_3$ by weight. ITO has been widely adopted by the solar industry as a transparent electrode due to both its high optical transmittance yet high electrical conductivity. In addition to the photovoltaic applications, new possibilities for ITO are emerging for electro-optic modulation. However, ITO has been found to be a difficult material to work with due to its processing invariances. In this review we first focus on ITOs fundamental properties, and in a second step relate them to the device details of high-performance electro-optic modulators.

### 3.1 Indium Oxide and Tin doped Indium Oxide Structure

The atomistic structure of a unit-cell of ITO resembles that of Indium Oxide ($In_2O_3$), with the latter being an ionically bound semiconducting oxide, which crystallizes in a cubic bixbyite-type structure with a space group Ia3 and a lattice constant of 10.118 nm [40]. The unit cell of ITO contains 80 atoms, and the Indium cations are located in two different six-fold-coordinated sites. One fourth of the cations are located in trigonally compressed octahedral, referred as *b* sites (In-O distance: 2.18Å), while the remaining three quarters are located on highly distorted octahedral *d* sites (set of three In-O distances: 2.13, 2.19 and 2.23Å) [41]. As Fig. 1a&b, indium atoms, on both *b* and *d* sites, reside at the center of a distorted cube with the six corners occupied by oxygen atoms, while the remaining two corners are empty. In the case of the *b* sites, oxygen vacancies are located along the body diagonal; for the *d* site they are located along a face diagonal [41, 42]. Figure 1c shows Tin (Sn) doping sites in an indium oxide lattice, where the Sn atom occupies an interstitial site and contributes an electron (i.e., a donor), making the doped indium oxide into indium tin oxide.

### 3.2 Indium Oxide and ITO Band Structure

The actual material properties of ITO have been found to be highly sensitive to the processing conditions [43]. While the detailed insights of band structure, as a function of processing parameters is desirable, the band structure is a function of the complex unit cell lattice and respective electronic interactions. Therefore, the properties are usually discussed in terms of an assumed band diagram consisting of an isotropic parabolic conduction band [41, 44]. A first attempt for its description was made by Fan and Goodenough [45] who proposed the schematic energy band model for indium oxide and ITO, which allows for a qualitative explanation of the observed high optical transparency and electrical conductivity in ITO (Fig. 1d). In Fan's band model indium oxide exhibits a wide direct band gap (3.5 eV) prohibiting interband transitions in the visible range and hence making it transparent within this frequency range, which is still the reference model used today for ITO. The conduction band was proposed to arise mainly from the Indium 5s orbitals and the valence band from oxygen 2p electrons. The Fermi energy $E_f$ is found a few eV's below the conduction band due to n-type doping of the Tin impurities [41]. The bands are slightly different; for low doping density donor states are formed just below the conduction band, and the Fermi energy lies between the donor level and the conduction band minimum (Fig. 1d). However, for very high doping, the donor density increases. Note that the donor states merge with the conduction band at a certain "critical" density, $n_c$, known as the Mott density defined by $\frac{n_c^{1/3}}{a_0^*} = 0.25$ for ITO, where the effective Bohr radius $a_0^*$ is about 1.3 nm for $In_2O_3$ resulting in $n_c$(ITO)=3.43×10$^{-19}$cm$^{-3}$ [41]. Above the critical Mott density, the impurities and the associated electrons occupy the bottom of the conduction band forming a degenerate electron gas, and the electron gas increases the ionized-impurity scattering and reduces the mobility [46, 47].

### 3.3 Optical and Electrical Properties of ITO

In order to utilize ITO for nanophotonic devices, a detailed knowledge and parameter interdependencies relating to the optical and electrical properties must be known. In general, the optical properties were previously modeled by a Drude-Lorentz model, due to the high carrier concentration as we discuss further below [49]. Focusing here on the electrical properties, the carrier concentration and mobility of ITO have been found to vary widely with process conditions for the deposition method such as sputtering power, oxygen flow, and annealing temperature. We therefore limit the discussion to a few key observations based on ITO's resistivity ($\rho = \frac{1}{ne\mu}$, where $n$ is the carrier concentration, $\mu$ is the mobility and $e$ is the free electron charge). For ITO, electrons are the majority carriers originating mainly from the doping donor Sn and oxygen vacancies. Therefore the carrier concentration of ITO films depends on the oxygen vacancy concentration, and the Sn doping levels. The latter is important since substituting an Indium atom frees an electron increasing the carrier density. However, doubly-charged oxygen vacancies and singly-charged Sn on an Indium site reduce the mobility of charge carriers due to ionized-impurity scattering [41]. Based on this understanding of ITO's doping interplay, the resistivity for ITO films can be tuned, either to a minimum such as for photovoltaic applications, or to a high-resistivity level. The latter is significant for electro-optic modulation as is discussed in section 4 in more detail.

Matrion et al. provided a useful, yet qualitative, overview linking the apparent as-deposited color of ITO films to the films resistivity, and the $O_2$/Sn doping concentration (Fig. 2a) [48]. The observed resistivity minimum (i.e. 'well') depending on both the $O_2$ and Sn doping levels, does explain the empirically observed challenge to tune ITO's resistivity; that is, the resistivity is high for low doping concentrations due to low carrier concentration, and low for high doping due to scattering from oxygen vacancies which reduce the mobility. Moreover, the observed apparent film illustrates that the higher doping could induce a particular state "extinguishing" Sn, but the high oxygen flow could not be well related to the color on the right side [41]. Interestingly, another group found the other resistivity-well trend as a function of sputtering power density (Fig. 2b) [50]. Increasing the power density creates ionized impurity centers, which induce scattering of the free electron carriers and then reduce the mobility. The gradual increase in the carrier concentration between a power density of 1 and 3 $W/cm^2$ is closely associated with the increased number of substituted Sn ions in the lattice. Increasing the sputtering power density further results in a flat carrier concentration due to a cancelation effect, which is from the increased level of Sn ion substitution and the decreased number of oxygen vacancies from the increased ITO film crystallinity. Many groups reported strong resistivity decreases with higher annealing temperatures; note that the annealing temperature leads to the related physical behavior as for the power density, as shown by the similar curve trends (compare Fig. 2b&c). However, the outlier data point (i.e. last data point for carrier concentration in Fig. 2c) actually drops instead of keeping constant, which may possibly be caused by the different oxygen flows used. Lastly, tuning with the oxygen flow the resistivity-well effect is present as well (Fig. 2d). Higher carrier densities are expected with comparatively low oxygen content and vice versa, and high mobility with increasing oxygen content, because of the enhanced crystallization. Since both mechanisms cause the opposite effects, the specific resistivity shows a minimum in the oxygen flow level. In summary these results show explainable trends for the resistivity, and color of ITO films as a function of doping, oxygen flow, deposition power, and substrate temperature. However, it seems the error bars are still rather large, while the more quantitative data and processing details are needed towards guaranteeing a to-spec engineered ITO film electrical property.

### 3.4 Typical Deposition Methods

This section highlights a variety of typically deployed methods for the deposition of ITO films. The different deposition conditions for each method significantly change the optical and electrical properties of ITO film.

#### *3.4.1 Sputtering*

Sputtering is one of the most widely used techniques for the deposition of TCO films. Within sputtering, a variety of techniques have been adopted including conventional direct current (DC), radio frequency (RF), and magnetron reactive sputtering of a metal alloy target in the presence of oxygen, as well as RF and ion beam sputtering of a pressed oxide powder target [41]. Among these options, DC and RF magnetron sputtering are the most attractive techniques for industrial development because both allow for high deposition rates, good reproducibility, and the possibility of using commercially available large area sputtering systems [53]. Typically, magnetron-sputtering processes are performed at high substrate temperatures ($\geq$ 200 °C), as these allow the best results in terms of layer transparency and conductivity to be obtained. However, several applications, for example solar cells and devices on plastics, require a low deposition temperature, as higher temperatures would damage the underlying either electronic device structures or substrate itself. This challenging task has been investigated by several workers where the RF technique was mostly adopted [54, 55]. The main difficulty of RF deposition at room temperature is due to insufficient recrystallization of ITO at low temperatures leading to the poor structural and electrical properties [56]. Furthermore, DC magnetron sputtering at room temperature is also challenging since the formed ITO layers are usually amorphous with a high electrical resistivity [57]. High temperature ITO sputtering can be done in DC and RF discharge plasma modes which allowing layers with obtainable similar transparency/conductivity properties. Depending on the plasma mode (DC or RF), an essential difference of optimum sputtering conditions (i.e. discharge power density, pressure, oxygen concentration, etc.) was observed, as well as a difference in the crystalline structure and morphology of the formed layers [53]. In general, sputtering allows for a relatively wide range of film tunablility due to the various process conditions involved in the process. However, the very process flexibility also requires precise control of all parameters.

#### *3.4.2 Evaporation*

For evaporation methods such as electron beam evaporation and direct thermal evaporation, typical powders of metallic Sn and In are commonly used. Regarding the latter, Indium will be preferentially vaporized, causing a change in the composition of the tin and indium liquid alloy with time due to the significantly higher vapor pressure of In relative to Sn [58]. That being the case, an electron beam can be a more reproducible method for low resistivity and high transparency films. Furthermore, low resistivity ITO films processed at low temperature have been reported using reactive electron beam evaporation onto heated glass substrates (300 °C and an oxygen pressure above 0.5 mT) [59]. Under optimum conditions, the resistivity is $3\times10^{-4}$ Ω·cm with an average mean transmittance of about 84% in the visible range [41].

*3.4.3 Chemical Vapor Deposition*

Chemical vapor deposition (CVD) is a process in which a chemical reaction involving gaseous reacting species takes place on, or in the vicinity of, a heated substrate surface. This technique has the advantage of being cost effective with respect to the apparatus, and enables the production of coatings with good properties without the use of high vacuum even on substrates with complicated shapes. The deposition of ITO films by the CVD method generally faces difficulties due to a lack of volatile and thermally stable source materials [41]. Under certain conditions, an ITO film with a resistivity as low as $2.9\times10^{-4}$Ω·cm is obtained at a reaction temperature of 400 °C [41, however more research is needed in this field.

## 4. ITO for Electro-optic Modulation

ITO has recently sparked the interest for active photonic components on-chip. In particular, the research by the Atwater, Leuthold, and Sorger group had raised the interest for electro-optic (EO) modulation utilizing ITO as an active switching material [60, 61]. For instance Sorger et al. demonstrated a 3-λ compact EO modulator using ITO inside a plasmonic hybrid mode [20]. In this section we summarize these effects along with some original content on ITO processing for high-efficient EO modulation. Furthermore, we review biasing options for ITO to produce epsilon-near-zero characteristics for EO modulation, and close this section by critically reviewing some outstanding questions in the field of ITO for EO modulation that shall guide the future research.

### 4.1 Drude-Model and Refractive Index Ellipsometry

The aim for strong EO modulation is to alter the effective modal index of an optical mode traveling inside the device. Note, in this discussion we exclude localized modulation effects, i.e. non-propagating, modes towards maximizing the index change of the active material, $\Delta n_{\text{active}}$, while minimizing the utilized voltage bias. Since the latter is related to the electrostatics inside the EOM, this discussion is postponed to the section 5, where a variety of experimental demonstrations of ITO for next generation EO modulation are highlighted and discussed. In what follows now, we discuss the possible strength of ITO-based EO modulation, which connects ITO's index to the carrier concentration via a Drude-Lorentz model. This model has been previously used for transparent conductive oxides such as ITO [62, 63], and allows us to write the complex permittivity as

$$\varepsilon(\omega) = \varepsilon_\infty - \frac{\omega_p^2}{\omega(\omega+i\gamma)}\ ;\ \omega_p^2 = \frac{n_c e^2}{\varepsilon_0 m^*} \qquad (1)$$

where $\gamma$ is the electron scattering rate, $\omega$ is the angular momentum in rad/s, $\varepsilon_0$ and $\varepsilon_\infty = 1+\chi = 3.9$ [35] are the free space and long-angular-momentum-limit permittivities, respectively, $\omega_p$ is the plasma frequency, $m^*$ is the reduced mass of ITO equating to $0.35\ m_0$, with $m_0$ being the rest mass of the electron, $n_c$ is the voltage-modulated carrier density for the ITO film, and $e$ is the elementary electron charge. Using Eqn. (1) and sweeping the free parameter, $n_c$, versus the device operating frequency, the general function for the permittivity and index are obtained (Fig. 3). The Drude-Lorentz model assumes the physical EO modulation origin to be the free carrier modulation. Thus, if the carrier concentration in a waveguide-type device design can be electrically altered, then EO modulation can be expected. However, only meaningfully strong modulation can be achieved if the ITO's index change may influence the effective mode index of the device, and not just the material films index. This therefore requires the mode field enhancement techniques such as observed in plasmonics, slot-waveguides, or optical cavities. The underlying physical mechanism is to enhance the non-linear polarization of the active material leading to a permittivity and hence index change as a function of applied voltage bias. This can be done by confining the mode into sub-diffraction limited fields, which increases the field strength at the active material, and thus leads to higher modulation efficiency (i.e. extinction-ratio/$V_b$). It is remarkable to notice that if the carrier concentration of ITO can be changed from ~$10^{19}$ to ~$10^{21}$ $cm^{-3}$, the index changes dramatically; the real part by unity, and the extinction coefficient by about 2-orders of magnitude (Fig. 3). Since the electrostatics of a device limits the achievable change of the carrier concentration, $\Delta n_{ITO}$, it is imperative to process ITO in such a way that the as-deposited film pertains carrier concentration as low as possible. This is important, because a given finite voltage bias will increases the carrier density by only a finite amount dictated by the electrostatics in an accumulation layer formed in an MOS capacitor. For instance the experimental results of Ref. [20] indicate an about 70 times carrier density increase for ITO when being sandwiched in a plasmonic MOS design. Note, that the optimization goal of processing ITO to exhibit a high-resistivity for the as-deposited films is opposite to the material goals of the photovoltaic industry, which utilizes ITO as a transparent electrode, where the aim is to decrease resistivity. However, EOM devices face a double-sided challenge when it comes to ITO's resistivity; when high-speed modulation is desired requiring low RC-delays, the contact resistance to ITO should be minimized. Thus, the carrier density seems to be a trade-off of the device performance, thus balancing signal extinction ratio (ER) and low energy consumption (i.e. energy/bit) on one side, and modulation speed on the other. However, this dilemma can be bypassed via spatially selective ITO plasma treatments which n-dopes ITO in a selected region at the electrical contact (i.e. assumed ITO is being electrically contacted serving as the active material). The latter case is discussed in more detail below. The outstanding aim for an ITO-based EOM optimization is therefore to achieve low carrier concentrations of ITO during processing for the device region.

We performed a variety of experimental ITO processing tests on for EO modulation, choosing the physical vapor deposition (PVD) method of sputtering to investigate effects of (a) $O_2$ concentration, (b) RF Power, (c) bias effect such as DC vs. AC, and (d) the effect of a wafer substrate. Regarding the results, we report on two selected parameters affecting the ITO's refractive index most strongly: the oxygen flow and the RF bias, whose effect can be visually seen by the optical appearance of the as-deposited ITO film (Fig. 4a), and by ellipsometry refractive index data (Fig. 4c). At a wavelength of 1550 nm, an increase in the $O_2$ chamber concentration clearly increase the refractive index, which correspond to a lower carrier concentration. It is consistent with the result of section 3, where a higher oxygen flow prohibits the existence of oxygen vacancies. However, altering the RF bias voltage induces a minor carrier concentration decrease. Note, that similar ITO film properties were observed between electron-beam evaporated films and sputtered films without oxygen flow, which can be explained by the lack of $O_2$ in both cases.

It is worthwhile to report our procedure and details of the performed ellipsometry tests. Variable angle spectroscopic ellipsometry (VASE) is a rather complex procedure and many physically non-meaningful results can be generated. For ITO the procedure is specifically challenging for several reasons due to (a) low signal-to-noise ratio owing to optical absorption, (b) strong dispersion (i.e. for broadband ellipsometry), and (c) vertical index variations since ITO grows commonly in graded microstructures. Following Ref. [66] we are able to provide the reliable ITO results following a 2-step procedure; ITO films are first fit with a Cauchy model over a smaller and typically less optical lossy wavelength range (i.e. 300-600 nm) to determine the film thickness. In a second step the Lorentz Drude model is used to obtain the broadband index information using the film thickness as an input parameter. Depending on the film thickness and application a graded index profile for ITO can be assumed for more accurate results. This relates to a gradually changed index profile with the film thickness due to the formation of an accumulation layer in the EOM devices. However, even following this procedure blindly is not advised, but a second control mechanism is needed for accurate results. That is, both angular functions Psi & Delta are fitted by both a material and a layer-composition model returning a residual mean error (RME) value (Fig. 4b). We found that the RME values above 10 are to be discarded, with our results reported here having RME $< 5$. If higher accuracy is desired the fitting procedure should be repeated until an RME minimum is found. We also tested the Tauc-Lorentz model for the step-1 fit, which provided the relatively stable results. As for the data presented here, an increase in oxygen flow shifts ITO to drop its loss, in accordance to the Kramers-Kronig (KK) relations, increase its refractive index (real part). A higher sputtering RF bias voltage has a similar effect, however with less severity than oxygen.

Of particular interest for EOMs is a material's ability to change its refractive index as a function of voltage bias. Early tests by Feigenbaum et al. performing ellipsometry on metal-oxide-ITO-metal heterostructures left the field with open questions [60]; the group argued that a capacitive accumulation layer is formed at the ITO-$SiO_2$ interface leading to the ITOs index modulation (Fig. 8 below). However, these data showed a decreasing loss with increasing carrier concentration, which appears non-physical or at least, unlike trends found in semiconductors. Moreover, as the loss decreases with bias, the real part was found to decrease simultaneously which seemingly violates the KK relations. It is for these non-congruent findings, that we repeated the biased ITO VASE experiments following the above-described methodology. In addition these measurements are extended into the telecommunication frequency range. Unlike Ref. [60] we find an increase of the ITO's loss with a bias voltage, which denotes an increase in optical loss due to carrier accumulation at the ITO-oxide interface (Fig. 5). Noteworthy is the sudden index jump with voltage and range; we note an increase from about 0.006 – 0.14 for the top 5 nm of a graded ITO film. Critical point for low RME is that the top metal acting as an electrode is semi-transparent. For the top metal contact a skin-depth thickness is selected for a wavelength in the NIR.

A band diagram analysis shows the band bending of the ITO layer at the ITO-$SiO_2$ interface (Fig. 5c). Note, ITO conduction band drops below the Fermi-level, resulting in an accumulation of electrons for positive voltage bias at the top-metal contact. Note, this is consistent with the measurements from Ref. [20], where no modulation effect was observed for a negative voltage bias. Furthermore, the electrostatic of the MOS capacitor is primarily governed by the 'gate' oxide layer (i.e. thickness and dielectric constant). Naturally, a stronger (weaker) electrostatics on the ITO layer, i.e. thinner (thicker) $t_{gate\text{-}ox}$, leads to an earlier (delayed) turn-on of the optical loss and hence OFF-state of an electro-absorption modulator. On the other hand a thinner $t_{gate\text{-}ox}$ also leads to a larger EOM device capacitance slowing down the 3-dB bandwidth. It is therefore interesting to note, that $t_{gate\text{-}ox}$ is a trade-off EOM device parameter that should be optimized for a specific EOM performance (i.e. low power, or high speed).

### 4.2 ITO Selective Contacts

Above we argued that in order to achieve a high modulation depth (i.e. extinction ratio) for EOMs, ITO should be deposited for a low-loss state (i.e. ON-state in EOM terminology), because raising the carrier density via a voltage bias increases optical loss shifting the EOM into the absorption (OFF) state. However, if ITO is to be electrically contacted, this poses a dilemma since ITO's low-loss state comes with low carrier concentration, and hence high resistivity specifically for thin films below 50 nm thicknesss. The latter makes high-modulation speeds challenging due to the limited RC delay time of the high resistive contacts. A potential solution explored here is to deposit ITO in a low-loss state at the device area, but then selectively treat ITO at the contact area. Since micrometer-scale local annealing is challenging, we explored selective plasma treatment. This method exposes ITO to an $O_2$ plasma thus doping the contact. Figure 6 summarizes 4-probe resistivity results for the various process conditions. The EOM optimized deposition method yields the expected high resistivity, but annealing and more importantly an exposure to $O_2$ plasma reduces the resistivity by several orders of magnitude. While these results are encouraging, a more in-depth study should be carried out to map-out this effect in detail, and we invite readers to perform this study.

### 4.3 Epsilon Near Zero ITO for Modulation

Recently the field of ultra-compact EO modulation has turned some attention to the concept of epsilon-near-zero materials, where the real part of permittivity is electrically tuned based on the Drude model [37]. Using free carrier density tuning can lead to a regime where the real part of index approaches zero, which in turn results in a dramatic change in the optical property of the material. Figure 3a above showed the carrier concentration condition where the ENZ effect occurs in ITO, which happens around a carrier concentration of about $10^{21}$ cm$^{-3}$ around a telecom frequency. It is interesting to note, that the experimental results of Ref. [20] where very close to this bias point, and hence are the first experimental evidence of this effect (even though the authors did not label it as an ENZ effect). The detailed picture is that ENZ materials impose a unique boundary condition onto electric fields (i.e. of nearby waveguides). This is a direct consequence of the continuity condition of the normal component of the electric field displacement, and leads in turn to large field enhancements inside the ENZ material relative to the adjacent layer. It is physical property of TCOs and hence ITO that makes them attractive for EO modulation applications. Recently, a few groups have explored this ENZ effect numerically; for instance Z.L. Lu et al. numerically reported an electro-absorption modulator based on tunable aluminum-doped zinc oxide (AZO) materials and slot waveguides [38]. The AZO layer employed as the active slot serves as an ENZ material, which can be tuned between ENZ (high absorption) and epsilon-far-from-zero (low absorption) by accumulation carriers. A.P. Vasudev et al. proposed a similar MOS-capacitor based modulator consists of a nanowire SOI waveguide coated with layers of 5 nm $HfO_2$ and 10 nm ITO [37]. Babicheva et al. numerically and analytically explored a variety of TCO EOM designs introducing a figure of merit for absorption modulators defined as

$$\mathrm{FOM} = \frac{|\mathrm{Im}(k_{\mathrm{eff}})_{\mathrm{on}} - \mathrm{Im}(k_{\mathrm{eff}})_{\mathrm{off}}|}{\mathrm{Im}(k_{\mathrm{eff}})_{\mathrm{state}}} = \frac{ER}{\alpha_{min}} \qquad (2)$$

where the extinction ratio is defined by ER $= \frac{10\lg\left(\frac{P_{on}}{P_{off}}\right)}{L} = 8.68(\mathrm{Im}(k_{\mathrm{eff}})_{\mathrm{off}} - \mathrm{Im}(k_{\mathrm{eff}})_{\mathrm{on}})$ (Fig. 7b) [36]. Benchmarking a variety of ultra-compact EOMs with Eqn. (2) shows that theoretically very high-performance EOMs are possible. Note, that this comparison does mix the experimental demonstrations with the numerical proposals.

## 5. EOM Devices based on ITO

Based on the potential for active EO modulation using ITO discussed above, this section reviews some initial experimental work demonstrating EOMs. In 2010, the Atwater group reported a method for obtaining unity-order refractive index changes in the accumulation layer of a metal-oxide-semiconductor heterostructure with ITO as the active material [60]. This pioneering work demonstrated a local $\Delta n_{\mathrm{index}}$ of 1.5 in the ITO film using plasmonic mode confinement as shown in Fig. 8a. Under an applied field, the structure forms an accumulation layer at the dielectric-conducting oxide interface. This increased carrier concentration in the ITO lowers the permittivity due to the Drude model. Furthermore, ellipsometry was performed on the entire heterostructure, until self-consistency was achieved between the complex indices in the heterostructure and the respective permittivity for each individual material. The measurements indicated a large change in the refractive index between 25-75%, at the visible wavelengths in a 5 nm layer (Fig. 8b). For 2.5 V, at 500 nm wavelength the index change is $\Delta n_{\mathrm{index}} = 0.41$ and at a wavelength of 800 nm the index change is $\Delta n_{\mathrm{index}} = 1.39$ [60]. However, the work left some open and unanswered questions such as why were the Kramers-Kronig relations violated, and the physical picture of increasing optical loss with increasing carrier density (see above). Although this work did not fabricate an actual EOM, their pioneer work pointed out that modulation of the local refractive index within these structures, combined with the high mode confinement achievable in plasmonic waveguides, could produce large changes in the effective index of propagating plasmonic modes, which gave ways to the future SPP EOM study as we will discuss next.

ITO has been implemented in a metal-insulator-metal (MIM) waveguide structure to demonstrate a sub-wavelength plasmonic modulator (Fig. 9a), for which a five percent change in the average carrier density (from $9.25\times10^{20}$ to $9.7\times10^{20}$ cm$^{-3}$) is observed [61]. The structure supports loosely bound SPP at a telecommunication wavelength of 1.55 μm owing to a MIM waveguide mode. A similar structure (Fig. 9b) based on a silicon-waveguide-integrated multilayer stack was fabricated and characterized. The logarithmic extinction ratio regarding to power up to 0.02 dB was achieved, which is much smaller than their prediction that 1dB extinction ratio on 0.5 μm length. Moreover, there is always a trade-off between modulation depth and transmittance due to the high confinement achievable in the MIM waveguide and the high losses associated with both metal and ITO layers. However, the group showed the first AC modulation of ITO EOMs into the MHz range.

In 2012, a record-high extinction ratio of 1 dB/μm was demonstrated for a plasmonic modulator utilizing a metal-oxide-ITO stack on top of a silicon photonic waveguide (Fig. 10) [20]. Under an applied bias, the carrier concentration is changed from $1\times10^{19}$ $cm^{-3}$ to $6.8\times10^{20}$ $cm^{-3}$, and the propagation length is varied from 1.3 to 43 μm. Also, the total insertion loss comprising of both the SOI-to-MOS coupling (-0.25 dB/coupler) and plasmonic MOS mode propagation (-0.14 dB/μm) were fairly low, which could achieve a total insertion loss as low as only about -1 dB for a 5 μm long modulator due to the good impedance match between the SOI and MOS mode and low ohmic losses from the plasmonic MOS mode [20, 67-69].

Recently, another characteristic of the SPP EOM deploying ITO as an active material was demonstrated by the Leuthold's team (2014); the hysteresis of the gate current and the optical transmission displays characteristics of a resistive random access memory (RRAM) (Fig. 10a) [70]. Note, the design is similar to that of Sorger's work (2012), and hence the observed ER of 1.2 dB/μm is expected [20]. However, by gradually altering the voltage an IV hysteresis is observed which indicates the memory effect of the switch, and the authors' claim arises from the metal pillars forming from the top metal through the gate oxide. The device demonstrates repeatability as indicated by a series of 50 consecutive measurements (Fig. 10b). Although the physical nature of the resistive switching mechanism is not fully understood, this work shows a new direction to utilize ITO as an active material.

## 6. Cavity-Enhanced Modulator Performance Enhancement

Next we are going to discuss how the performance of an electro-optic modulator's modulation performance (i.e. extinction ratio can be enhanced by introducing a resonator cavity (Fig. 12). Unlike the popular micro-ring resonator cavity such as demonstrated by the Lipson group , here we aim to design an ultra-compact waveguide-integrated (in-line) cavity where the footprint of the device is not extended beyond the index modulator. The device is designed on the basis of a photonic-plasmonic hybrid mode in a Metal- Oxide-Semiconductor (MOS) configuration previously investigated [20,67-69] featuring a deep sub- confinement yet reasonably high propagation lengths. i.e. a high modulator function-to-parasitic ratio. The EO switching mechanism is based on a thin voltage carrier modulated Indium-Tin-Oxide (ITO) layer inside the MOS configuration. Our question is to what extend can the resonator enhance the obtainable modulation extinction ratio (ER) vs its insertion loss (IL). TO show this, we select an integrated, in-line cavity on this waveguide-based electro-absorption modulator (EAM) design.

The general device design consists of a SOI waveguide with an ITO-$SiO_2$-Au stack placed on top (Fig. 1). This configuration forms a Metal-Oxide-Semiconductor (MOS) capacitor featuring an accumulation layer at the ITO-$SiO_2$ interface upon applying a voltage bias between the Gold and Silicon, which is key to our modulator design. When forward biased the carrier density inside the ITO increases making it more metal like and the imaginary part of the refractive index, $\kappa_{ITO}$, increases from about $10^{-4}$ (no voltage) to 0.273 (Voltage applied) [20]. The plasmonic MOS mode enhanced field overlaps strongly with the oxide and ITO, when the applied voltage is off and on, respectively.

Next, we discuss, the EO modulator mechanism of cavity device; the cavity refractive index changes as a voltage is applied, thus altering the devices' optical mode index change and hence its resonant frequency. To effectively confine the optical mode in the cavity, the cavity length *L* can be derived from the wave vector quantization condition:

$$k_x^2 + k_y^2 + k_z^2 = n^2 k_0^2$$

the dimensions in the equation should be reduced to 2D as we use 2D model in the simulation,

$$k_x^2 + k_y^2 = n^2 k_0^2$$

which yields:

$$L = \frac{\sqrt{2}}{2} \times \frac{\lambda}{n}$$

where $L$ is the cavity length, λ is the operation wavelength and $n$ is the device mode refractive index.

The task is now to find those geometric values for the cavity stubs and length as to optimize the performance of cavity- based electro-optic modulator. By sweeping those three parameters we monitor the EAM modulating extinction ratio (ER) and insertion loss (IL), i.e. Voltage off case. Considering providing an easy comparison metric and performance measure for our device, we report ER/IL per nominal device length of 1 μm. The one-micrometer device size in the metric was chosen as to allow for extreme dense opto-electronic chip integration.

For the numerical simulations, we build and tested our model in a commercial FDTD solver (Comsol) for a fixed operation wavelength of 1310 nm, proper for telecommunication (Fig. 13b). Before simulating the cavity performance, we found the ER and IL for the case without a cavity; the results show a reasonable strong ER as compared to non- plasmonic devices, while keeping a low power budget in the off state (IL < 0.5 dB). We started with a medium step size for the geometrical values stated above as to map out the optimum performance. The data from Figure 13a-f show multiple interesting results; (i) extremely high-performance metrics, i.e. ER/IL-μm exceeding 50, can be achieved, (ii) for a longer cavity length a the stub widths can be made more narrow, (iii) the shorter the cavity length, the wider is the ‘sweet-spot’ region, and (iv) the stub depth is somewhat less sensitive. The latter can be explained by the fact that part of the highly confined optical mode resides in the flat ITO layer above the stubs.

Exploring the performance further (Table 1), we list best performance values for the various cavity lengths explicitly; the about 10-fold performance increase relative to a device lacking the cavity can be clearly seen. Highest performer is the device with the cavity length of 140 nm, stub width and depth of 300 nm and 230 nm, respectively. Bringing this value in to contrast with other excising technologies like demonstrated by leading industry with ER/IL-μm values around $10^{-3}$-$10^{-2}$ [71], the device discussed here is 3-4 orders of magnitude higher performing. Notice while the IL is somewhat low but constant, by adding the cavity the extinction ratio is enhanced by about 10 times. This can be explained due to an enhance light- matter-interaction of the optical mode with the loss-changed ITO layer.

Summarizing, by adding a waveguide-based in-line optical cavity to a plasmonic EOM enhances the switching performance (i.e. extinction ratio) of the device by about one order of magnitude, while keeping the insertion loss below the 0.5 dB value for the device. This results in 3-4 orders of magnitude higher device performance compared to non-plasmonic pursued designs. The physical basis for this is (i) a low-loss deep-subwavelength optical mode, (ii) a highly switchable active material, ITO, and (iii) low quality factor cavity enhancing the light-matter-interaction. Notice, the device size of 1 micrometer allows for electrical capacitance values in the atto-Farad range, thus ultra-fast modulation can be achieved due to a low RC-delay time. The seamless integration of these high performance EO switching nodes into low-cost SOI platforms paths a way for high functionality future integrated circuits.

## 7. Outlook and Conclusion

In addition to the experimental demonstrations briefly discussed above, there are a variety of theoretical proposals for TCO and/or ITO-based EOMs [37, 38, 39, 67]. Common to most of them is the usage of plasmonics to enhance the LMIs, resulting in small footprint and high-performance electro-optic properties.

Changing the discussion to highlight exemplary recent results of ITO and demonstrated modulators, in a previous study we analyzed some scaling vectors towards achieving atto-Joule per bit efficient modulators on-chip as well as some experimental demonstrations of novel plasmonic modulators with sub-fJ/bit efficiencies [72]. In this parametric study of placing different actively modulated materials into plasmonic versus photonic optical modes shows that emerging EO materials such as TCOs (such as ITO) and 2D materials (such as Graphene) overcompensate their small optical mode overlap with the active material by the very large (unity) index modulation [73]. In addition, we show that the metal used in plasmonic modulators can act as a low resistance contact to the device (i.e. no doping is needed). In initial experimental demonstration of a photon-plasmon-hybrid graphene-based electro-absorption modulator on silicon showed indeed a high switching efficiency of sub-1 V and showing a high 0.05 dB $V^{-1}$ $\mu m^{-1}$ modulation performance at <1V of bias or about 110 aJ/ bit [73]. Improving on this demonstration, a slot-based graphene modulator design in a push–pull gating configuration increases this performance to 2 dB $V^{-1}$ $\mu m^{-1}$ allowing the device to be just 770 nm short for 3 dB small signal modulation.

Indeed, it is interesting to see that the 'nano-optics becoming practical' [74] article still bears relevance today. In the context of PICs, TCO-based $\lambda$-size EOMs might be a logical next technological platform for hybrid networks on-chip e.g. [75,76]. Fortunately, there is much tunablity in ITO and devices can be optimized for both EAM and EOM (i.e. amplitude modulation, $\Delta\kappa$, or phase modulation ($\Delta n$), depending on the carrier concentration inside ITO; for example in ref [77] we showed how a low carrier concentration ($10^{20}cm^{-3}$) enables high-performance phase shifters such as demonstrated in [78] demonstrating the first GHz MZI-based ITO modulator. In contrast, a plasmonic EAM with a strong index changing mode but high loss is best optimized by targeting the $\Delta\kappa$ region coinciding with a high carrier concentration (6-7x$10^{20}cm^{-3}$) such experimentally demonstrate in ref [20]. A third alternative is the actually utilize the (elsewise) parasitic Kramers-Kronig relations, i.e. the simultaneous delta-$n$ and delta-$\kappa$ modulation in modulators; here a EAM can be augmented with a phase-modulating directional-coupler section, where the unwanted light is not only absorbed ($\Delta\kappa$) but also 'deflected away from' the downstream waveguide utilizing ($\Delta n$). Following this, we earlier demonstrated a high voltage-tuning ability of 0.75dB/V in an ITO-based silicon waveguide design [79]. This concept may also lead to high energy-bandwidth where the modulation coupling section is distinct from any resonator enhancement section, i.e. the long photon lifetime of a high-Q resonator providing efficient switching (low Voltage), is decoupled form a low-Q modulation resonator. As an example, see work form the Hu group [80]. However, the tradeoff is naturally the larger footprint for both cavities forming the EOM.

Regarding ITO modulator speed, the analysis is relatively straightforward; there are two fundamental speed limiting factors; a) the time it takes for the carriers to move to form an accumulation layer and b) the RC-delay of the device. The former is estimated by the drift-time given a gate bias and ITO mobility. Typical mobilities in our group are around 45-50 $cm^2/Vs$, which result in a 10's THz fast signal potential. Hence, in a real device the limiting factor becomes RC. More specifically, since the unity high ITO material modulation capability, the device length of the active contact can be <10micrometer short. Hence, the capacitance is around 1-20fF for most of our devices. Hence, the challenge remains to achieve a low resistance ($R$) to the device (Fig. 14); the resistance breaks down into several 'sections'. Fortunately, using a plasmonic mode, as argued above, allows for the top contact to be ohmic ($R$ = 5-10Ω). The remaining challenge then becomes to achieving a low resistance to ITO, where two options exist (see left and right of Figure 14); either the ITO is contacted and the end (butt coupled), or from the 'side' (along the waveguide). Using our experimentally routine values for ITO gives about 150Ω and about 10Ω for the side-coupled option. Interestingly, these two approaches allow for speeds in excess of 60GHz and 800GHz, respectively.

However, a variety of open questions are still outstanding in the field; (1) foremost, the debate of the nature of the index modulation of ITO is unsettled, and consists of two different viewpoints; one favours the carrier-based dispersion shift of ITO described by the Drude mode, whereas the second allocates the formation of metallic pillars that growth from the metal (typically Gold) contact through the electrical gate oxide [70]. Interestingly, the latter was reported in the supplementary information of the Feigenbaum's original work already. In this regard a detailed study correlating the device performance, with modelling and cross-sectional SEM images would be needed to solve this outstanding question on the modulation mechanism. (2) Based on our own ellipsometry work, we must conclude that this technique is very sensitive to the model type chosen and the fitting procedure. Thus, in order to make results comparable, detailed procedures to minimize the fitting errors should be followed. (3) The localized and selective doping ITO in a post-deposition process step to lower the contact resistance should be studied in more detail for future work. Another interesting direction could be to explore ITO-based beam-modulating devices but adding non-volatility for memristive networks and machine learning ASICs [81].

In conclusion transparent conductive oxides, and particularly ITO, have been found to be intriguing candidates for strong refractive index modulation. The physical effect seems to point towards a strong carrier modulation of the TCO when biased near the epsilon-near-zero point of ITO's permittivity. Early studies have demonstrated electro-optic modulation devices that utilize plasmonic field confinement and these TCOs and later studies have demonstrated 10's GHz fast modulation. The next big challenge is to demonstrate 100+GHz fast modulators and integrate this important ITO platform into foundry PDKs. A final note on the component-footprint density on-chip; a to-scale comparison of ITO modulators to the prevailing Silicon MZI shows a more than 1000x smaller footprint [82]. While this may not matter as much for optical transceiver technology where only a 'few' modulators are used per-chip, such compact technology could become a game-changer for emerging ASIC processors such as photonic-tensor core [83] accelerators where in a 256x256 Multiply-accumulate (MAC) array 65,000 devices are needed.

While this research field is still on going, those early demonstrations have shown the potential to deliver $\lambda$-size, ultra-low capacitance devices for future photonic integrated circuitry.

**Acknowledgements**
We acknowledge support from the Air Force Office of Scientific Research (AFOSR) under the award number FA9559-14-1-0215 and FA9559-14-1-0378.

## Figure Captions

**Figure 1** Indium Oxide structure. **a**, Eight oxygen atoms are situated within a compressed octahedra (b site) and have six equidistant oxygen atom neighbors at 2.18Å. **b**, Eight Indium atoms are situated at the corners of a highly distorted octahedron (d site). For these Indium atoms, there are three possible cation-oxygen distances: 2.13, 2.19 and 2.23 Å [42]. **c**, Sn doping sites in an $In_2O_3$ lattice [48]. **d** Schematic energy-band model for tin doped indium oxide, where $E_f$, $E_y$ and $E_c$ is the Fermi energy, the violation band energy and the conduction band energy respectively. Left: low doping level, right: high doping level [45].

**Figure 2** ITO electrical properties **a**, Resistivity well effect and the color change due to different doping and oxygen concentration [48]. **b**, Mobility, concentration and resistivity as a function of deposition power density [50]. **c**, Concentration and resistivity with increasing substrate temperature [51]. **d**, Resistivity change due to tuning oxygen flow [52].

**Figure 3** ITO optical film parameters. **a** & **b**, ITO Permittivity dispersion for various carrier concentrations. **c** & **d**, Index dispersion of the active materials ITO based on the Drude-Lorentz model for various carrier concentrations [64]. Arrows indicate switching range for ON (arrow start) and OFF (arrow tip), respectively [65]. $\lambda$ = 1310 nm vertical dashed line.

**Figure 4** Optical performance measurements of ITO samples using variable angle spectroscopic ellipsometry method. **a**, Photography of four ITO samples with a variety of experimental ITO processing tests (i.e. different combinations of oxygen flow and RF bias), showing different appearance of ITO films. $O_2$ flow in sccm, and RF bias in volts. **b**, The angular functions, Psi & Delta as a function of wavelength range, where Psi is the ratio of the amplitude diminutions and Delta is the phase difference induced by the reflection, both were measured by ellipsometry method. **c**, The measured results of refractive index ($n$ is the real part and κ is the imaginary part) as a function of wavelength range with different oxygen flow and RF bias.

**Figure 5** Dynamic VASE on ITO MOS structures to determine the optical loss change as a function of voltage bias. **a**, Setup schematic. Note, no significant change in the results were recorded when ITO or the doped silicon substrate was contacted. Oxide = $SiO_2$ (ALD), metal = Gold. ITO and Au were sputtered. **b**, ITO film extinction coefficient vs. bias voltage shows a strong modulation. Graded ITO film, results for top 5 nm. **c**, Band diagram of the ITO-MOS design identifying the ITO-$SiO_2$ interface for the charge accumulation region.

**Figure 6** Resistivity of ITO films and contact pads. **a**, Micrograph of a test structure for transmission line method (TMD) resistivity tests. Note, the ITO film covers the sidewalls of the SOI waveguide. Scale bar 50 μm. **b**, Effect of the sputtering ITO films and annealing. Annealing at T = 300 °C for 30 min. Plasma treatment, $O_2$ flow-rate ~ 30 sccm, exposure time = 5 min.

**Figure 7** **a**, Epsilon-near-zero (ENZ) condition for the ITO after Ref. [37]. Bringing the real part of the material close to zero can result in large modal index changes of nanostructured waveguides. **b**, Comparison table of ultra-compact EOMs based on various TCO materials, where ER is the extinction ratio, $\alpha_{min}$ is the minimum attenuation, and FOM is the figure of merit. [36].

**Figure 8** **a**, MOS structure with (dashed red) and without (solid green) charge accumulation [60]. From top to bottom are the structure schematics, band diagram, carrier concentration, permittivity and refractive index, respectively. **b**, The complex refractive index ($n$ for the real part and κ for the imaginary part) in a 5 nm accumulation layer, extracted from the ellipsometry data.

**Figure 9 a**, Schematic of the waveguide-integrated, silicon-based nanophotonic modulator [61]. The MOS design features a plasmonic optical mode, which concentrates the mode's electric field and allows for a good overlap with the active ITO layer. **b**, 3D schematic of the fabricated device and its lumped element model describing the low-pass characteristic of the device.

**Figure 10 a**, Schematic of the waveguide-integrated, silicon-based nanophotonic modulator [20]. **b**, Electric field density across the active fundamental MOS region of the modulator. The MOS design (with a gate oxide of about 30nm) features a plasmonic optical mode, which concentrates the mode's electric field and allows for a good overlap with the active ITO layer.

**Figure 11 a**, Cross section view of the plasmonic memristor [70]. The device comprises a typical memristor MIM layers that further serve as a plasmonic waveguide on top of a silicon plasmonic waveguide. **b**, Latching optical switch behavior for a 5 μm long device: 50 measurement cycles of the normalized optical transmission as a function of the set voltage showing a hysteresis and an extinction ratio of 6 dB.

**Figure 12.** Cavity-enhancement options for EO modulators. The obtainable ER is enhanced by utilizing both an index-inducing resonance shift and an absorption modulation.

**Figure 13. a-f**, show a FOM = ER/IL-μm for different cavity length from 100 nm to 200 nm with each step of 20 nm. For a fixed cavity length, stub width and depth have been swept to reach a better FOM. Highest performance is achieved with cavity length of 140 nm, with stub width and depth being 300 nm and 230 nm, respectively. The cavity- based EOM is more sensitive to stub width than stub depth as can be seen by comparison of the gradient. Therefore, we altered the stub width in smaller steps compared to the depth. $\lambda$ = 1310 nm, device size = 1 μm. **g**, Schematic of a cavity-based electro-optic modulator. A Fabry-Perot cavity is created by introducing two stubs into the waveguide-based plasmonic MOS mode. The index of the cavity is changed by the creating an accumulation layer inside the ITO, shifting its extinction coefficient. ER/IL/μm of cavity modulator.

**Figure 14.** ITO-based EOM modulator speed considerations; performing a fundamental speed-limit analysis shows that the carrier drift-time is not a limiting factor (see text), but the RC-delay is (or mor specifically R). Here two bias schemes for ITO exist (assuming the other contract is a plasmonic metal, hence R is very low), shown left and right. The middle section shows the 'top-gate and oxide' removed for clarity. Using our routinely achievable ITO resistivities and realistic device design values from ref [82] a speed in access of 50GHz for but-coupled ITO and over 500GHz for edge-coupled should be achievable.

**Table 1** Comparisons of a variety of active EO materials [36]. TCO's allow for strong optical modulation due to their ability to modulate the carrier density over several orders of magnitude electrically. However, typically optical losses are high.

| Mechanism | Active material | Advantage | Challenge |
|---|---|---|---|
| Carrier concentration change | Si | Fast, Low-loss | Weak response |
| | III-V | | |
| | Graphene | | |
| | TCO (ITO, GZO, AZO) | Fast, Significant response | Lossy |
| Phase transition | $VO_2$, Ga, $BiFeO_3$, $BaTiO_3$, etc. | Significant response | High energy per bit Low-speed (1μs) |
| Thermo | Polymers | | |

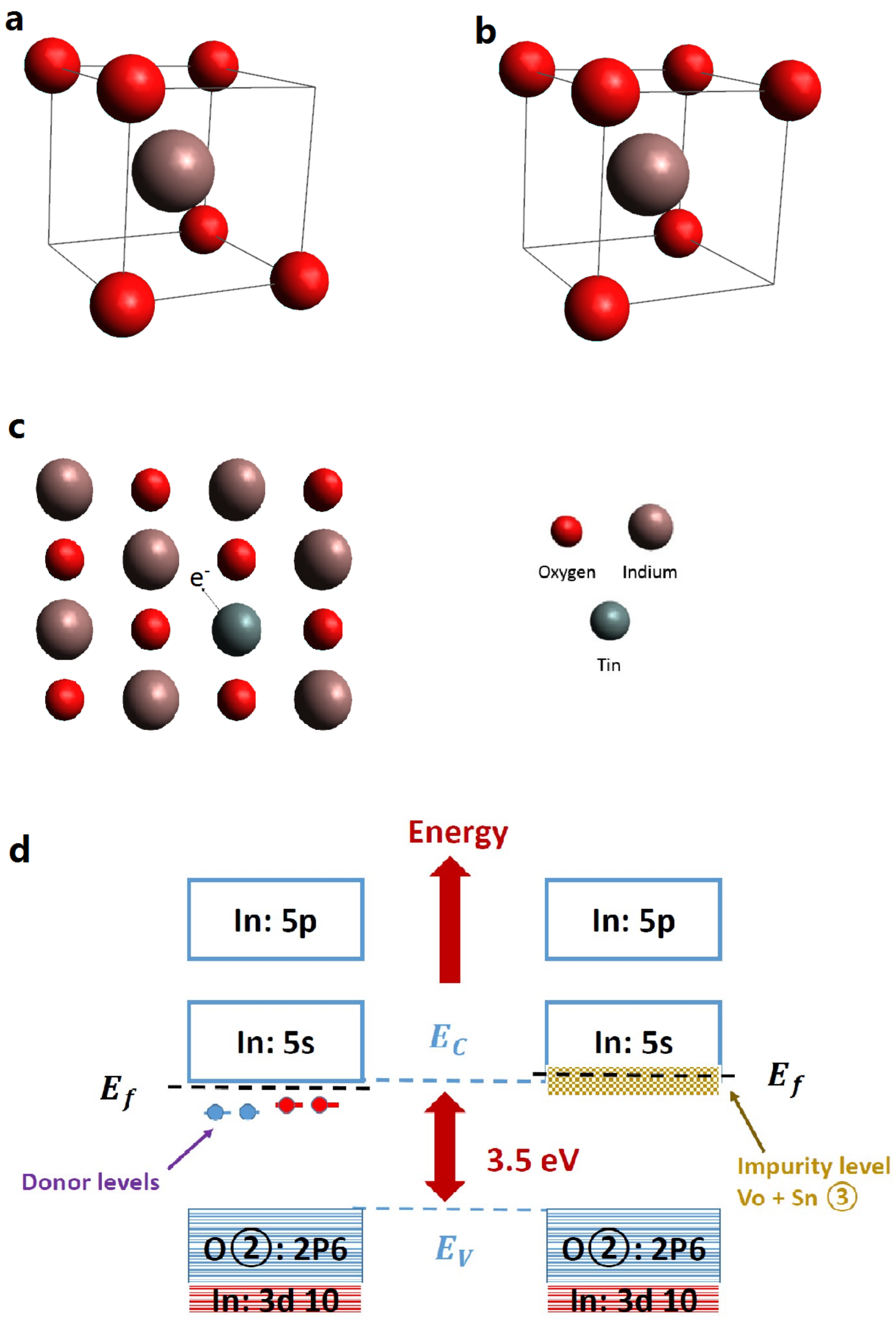

# Figure 1 of 14

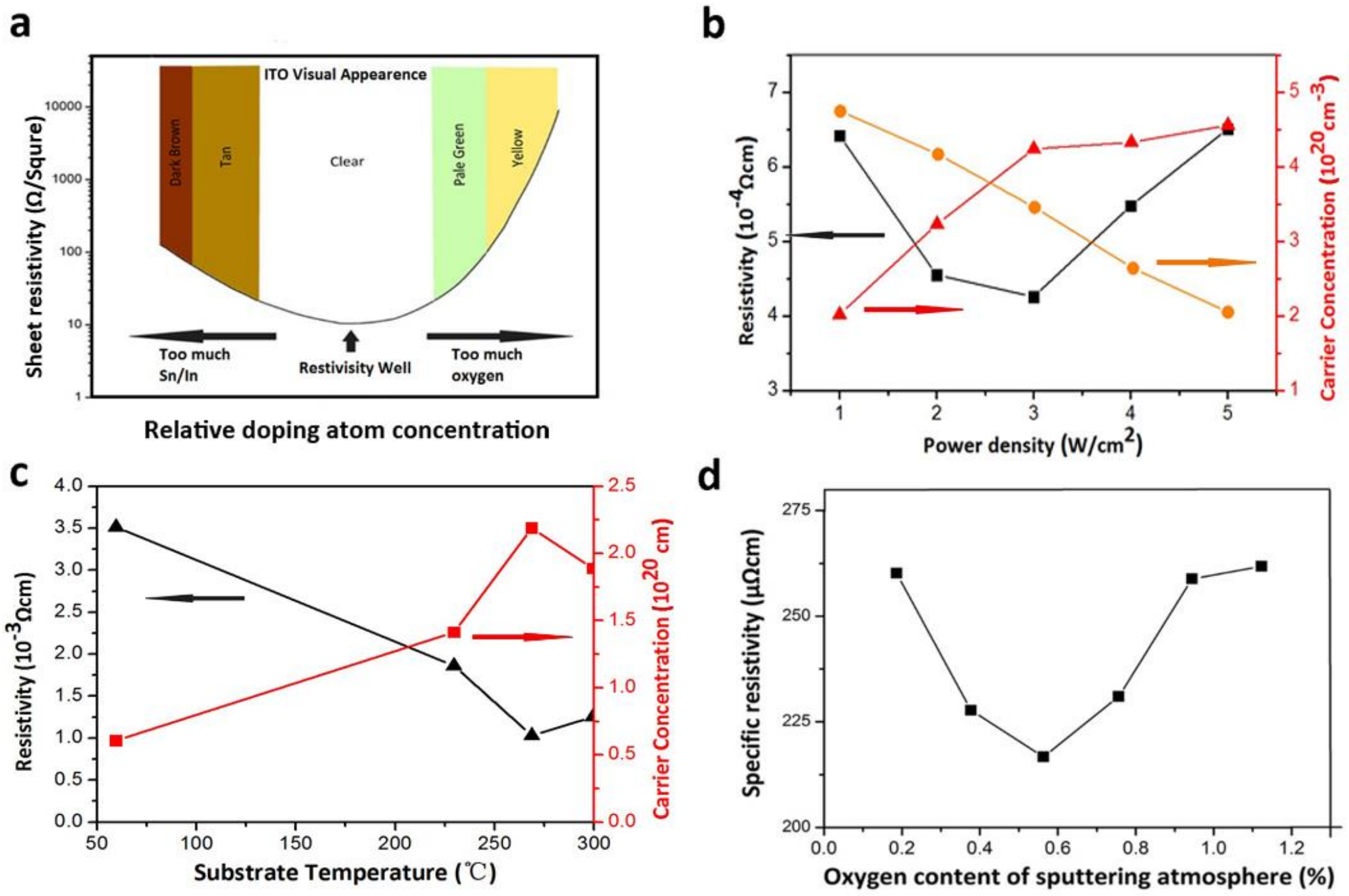

# Figure 2 of 14

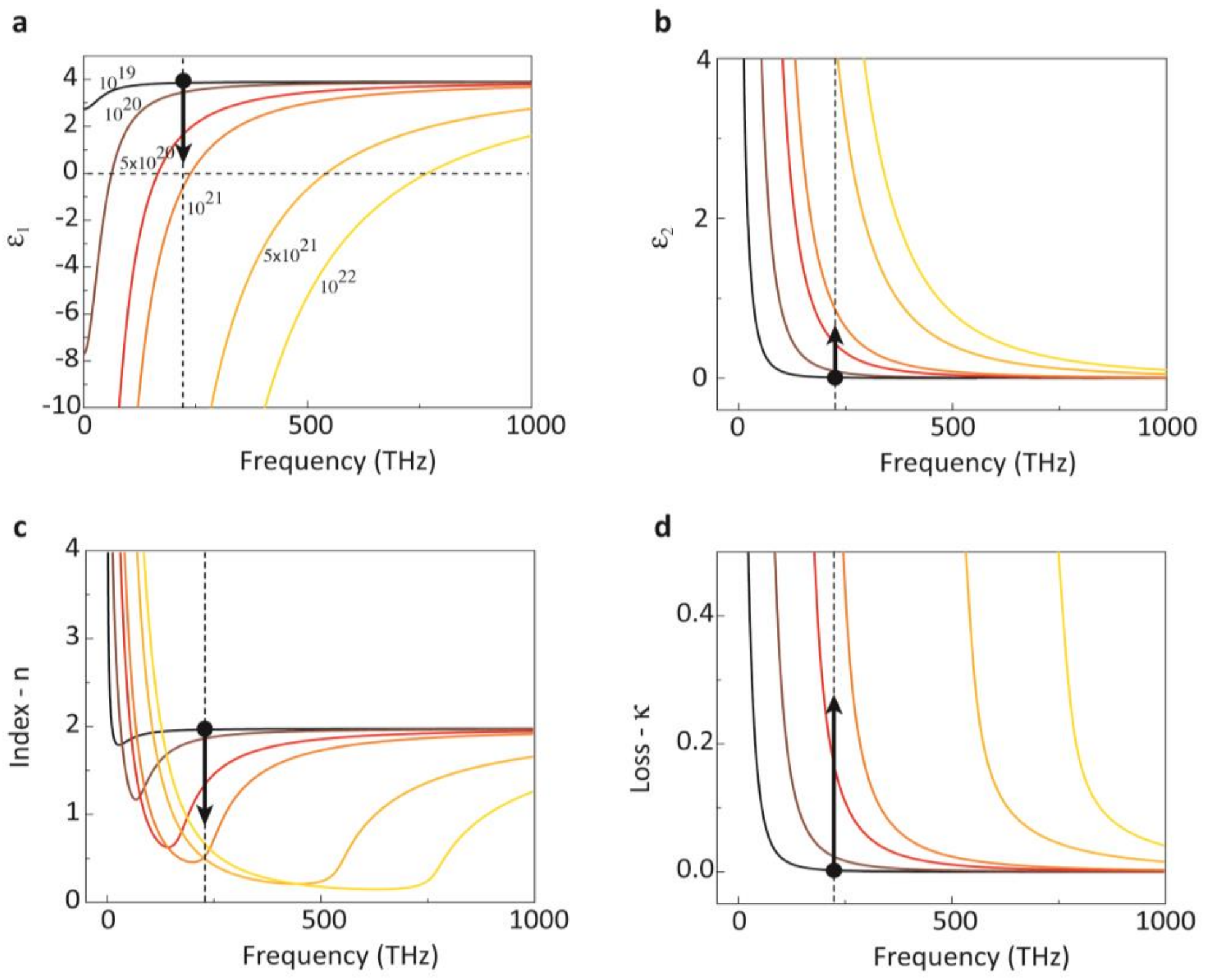

# Figure 3 of 14

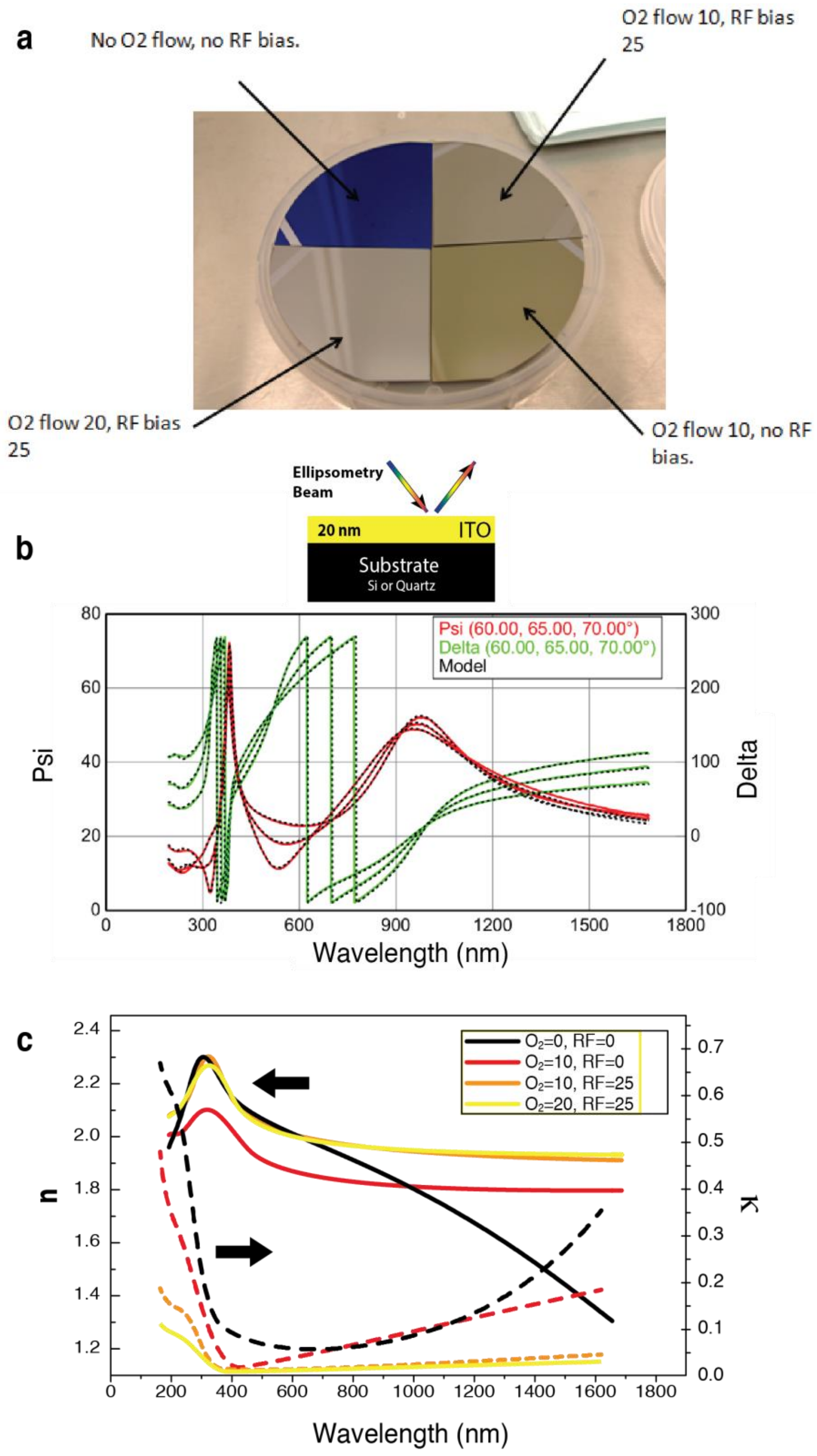

# Figure 4 of 14

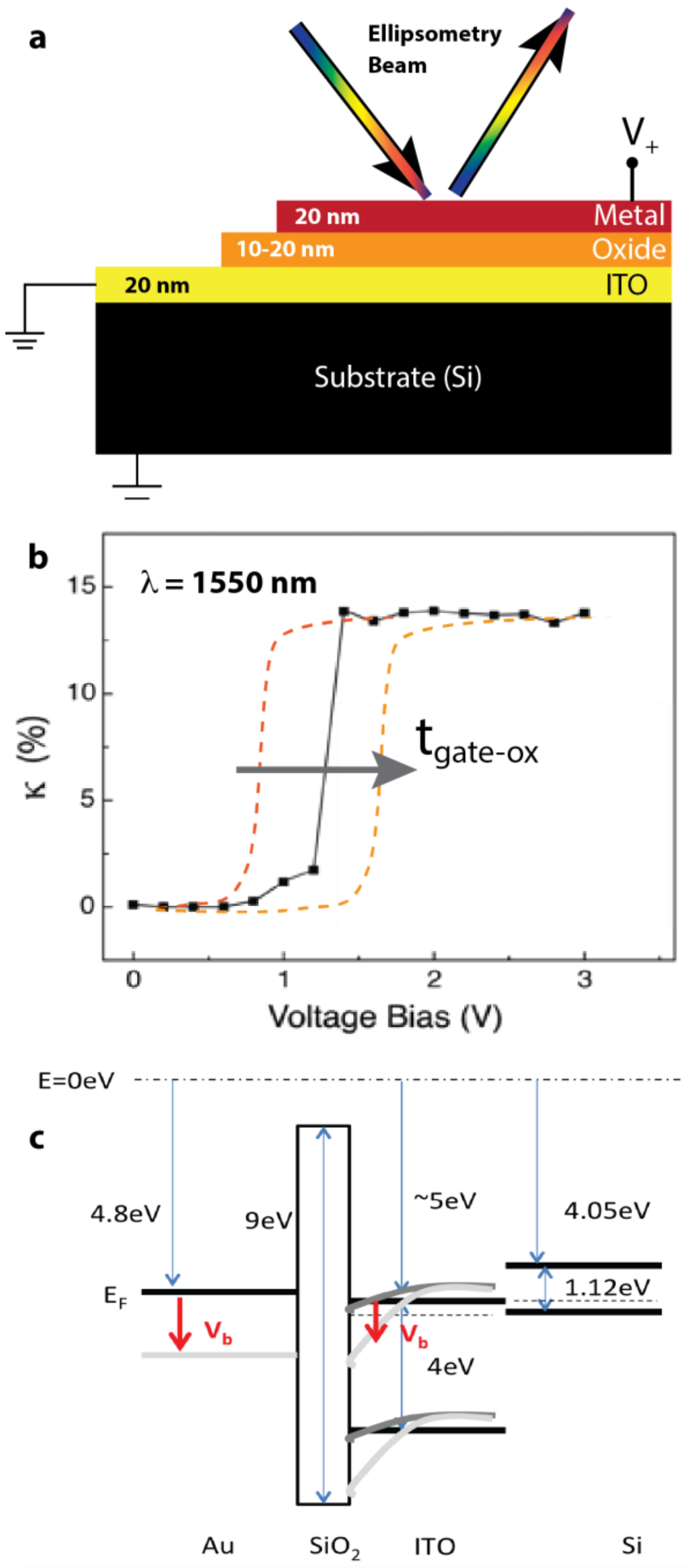

# Figure 5 of 14

**a**

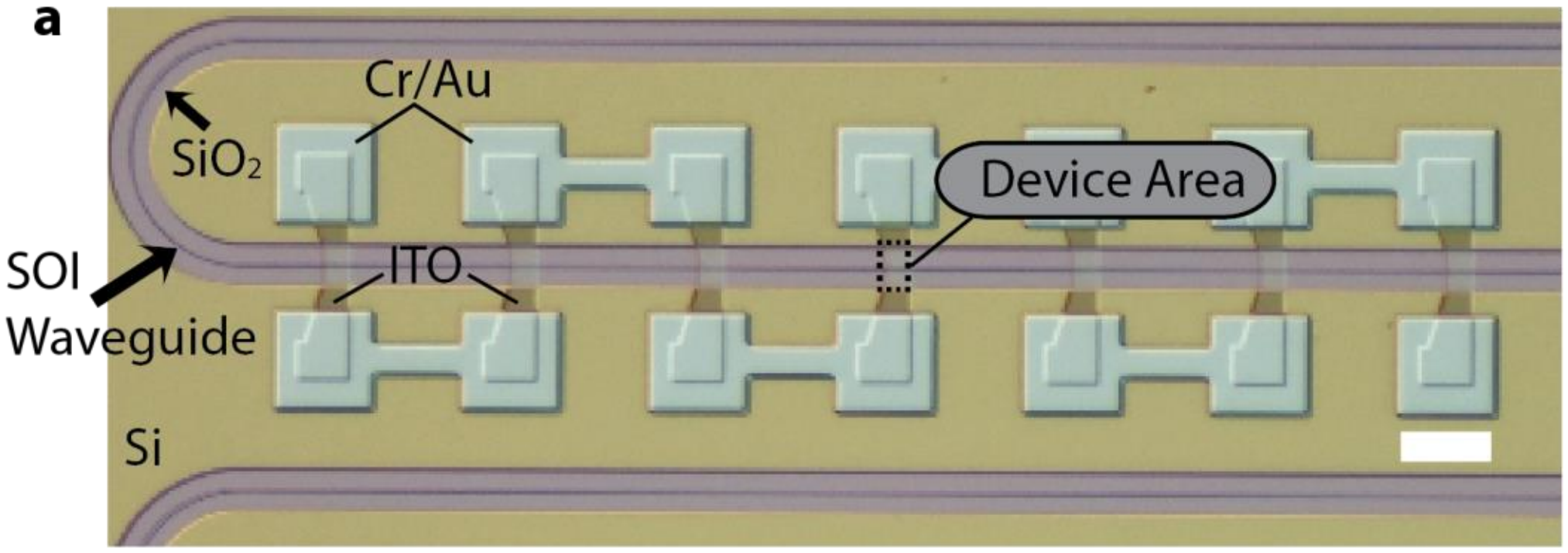

**b**

| Sample | Si/ITO $O_2$=20 sccm Annealed = no Plasma = no | Si/ITO $O_2$=20 sccm Annealed = 300C, Plasma = no | Si/ITO $O_2$=20 sccm Annealed = no Plasma = Yes | Si/ITO $O_2$=20 sccm Annealed = 300C Plasma = Yes |
|---|---|---|---|---|
| **Resistivity** [Ω-cm] | [$10^{-2}$-10] | ~$10^{-3}$ | [$10^{-4}$-$10^{-3}$] | ~$10^{-4}$ |

# Figure 6 of 14

**a**

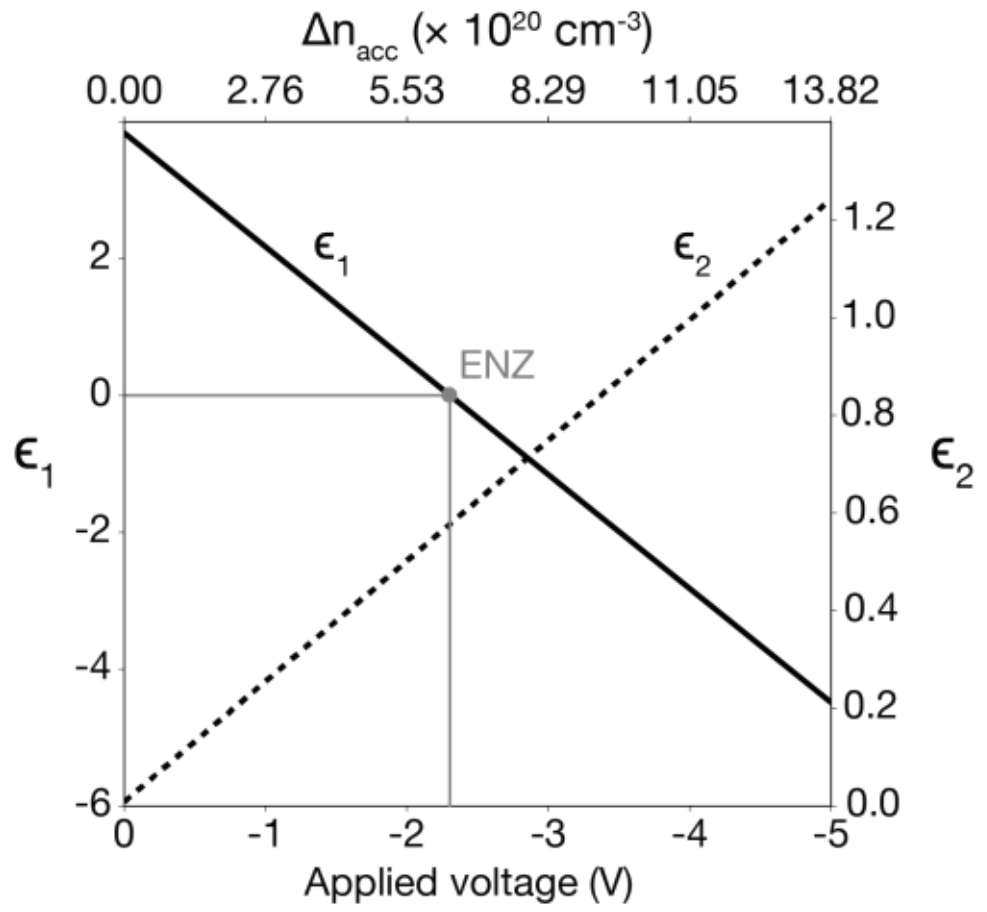

**b**

| Device | ER, dB/μm | $\alpha_{min}$, dB/μm | FoM | λ, nm | Based on | Date |
|---|---|---|---|---|---|---|
| SPPAM | 2 | 24 | 0.08 | 1550 | MIM | 04/2011 |
| 3λ-size | 1.0 | 0.04 | 25 | 1310 | Si-wg | 04/2012 |
| Tunable ENZ | 18 | 1 | 18 | 1310 | Si-wg | 06/2012 |
| Tuned ε | 3 | 9 | 0.3 | 1550 | MIM | 07/2012 |
| wire-MIM | 13 | 11 | 1.2 | 1550 | MIM | 09/2012 |
| Sub-λ-size | 6 | 0.7 | 8.5 | 1310 | Si-wg | 07/2013 |
| Slot | 2.5-11 | 0.85 | 3-13 | 1550 | Slot-wg | 10/2013 |
| ENZ-Si | 0.10 | 0.003 | 37 | 1550 | Si-wg | 11/2013 |
| GZO-Si | 24 | 0.06 | 400 | 1550 | Mult-wg | 11/2013 |
| TiN-Si | 46 | 0.3 | 160 | 1550 | Mult-wg | 11/2013 |

# Figure 7 of 14

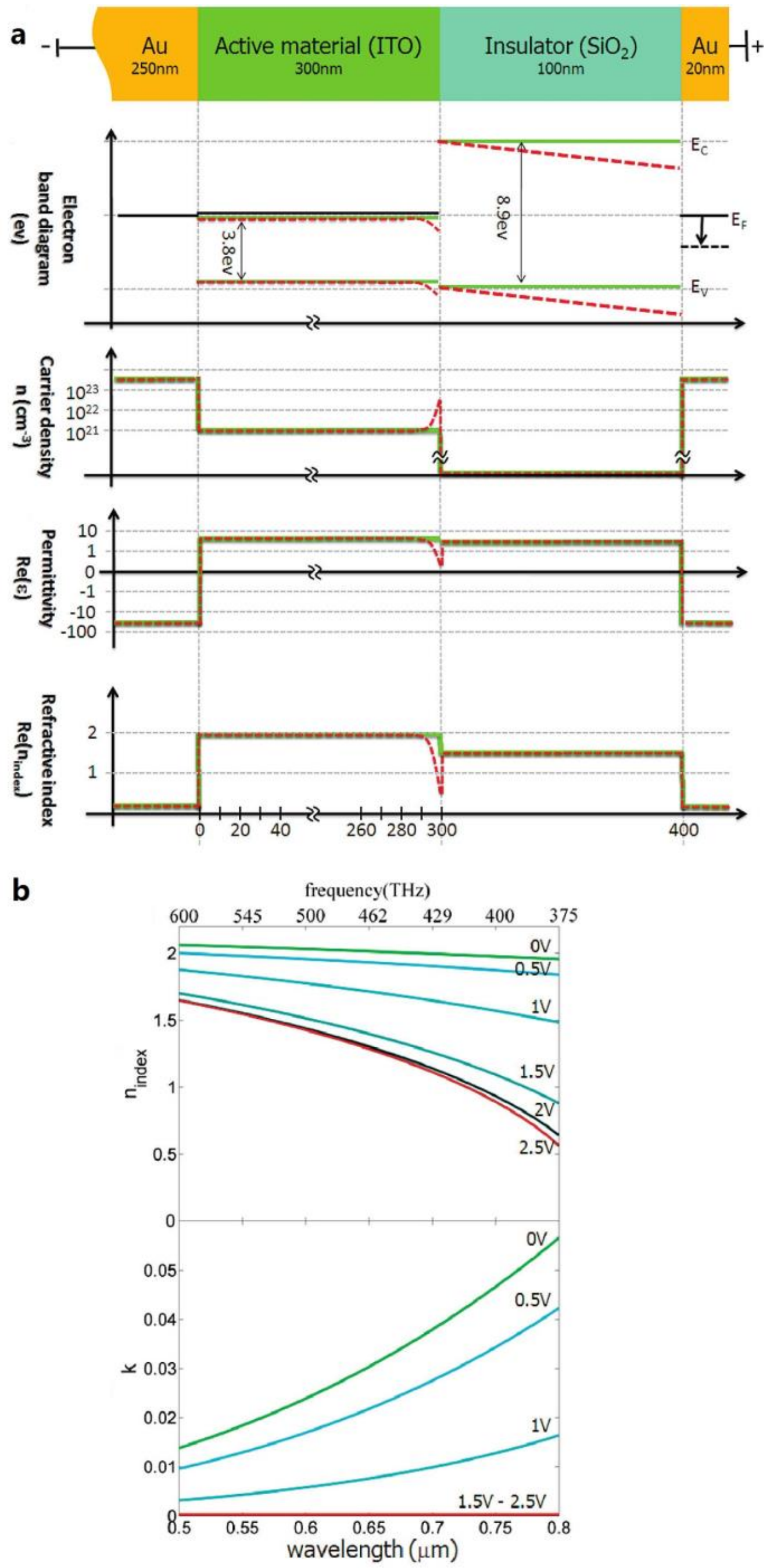

# Figure 8 of 14

**a**

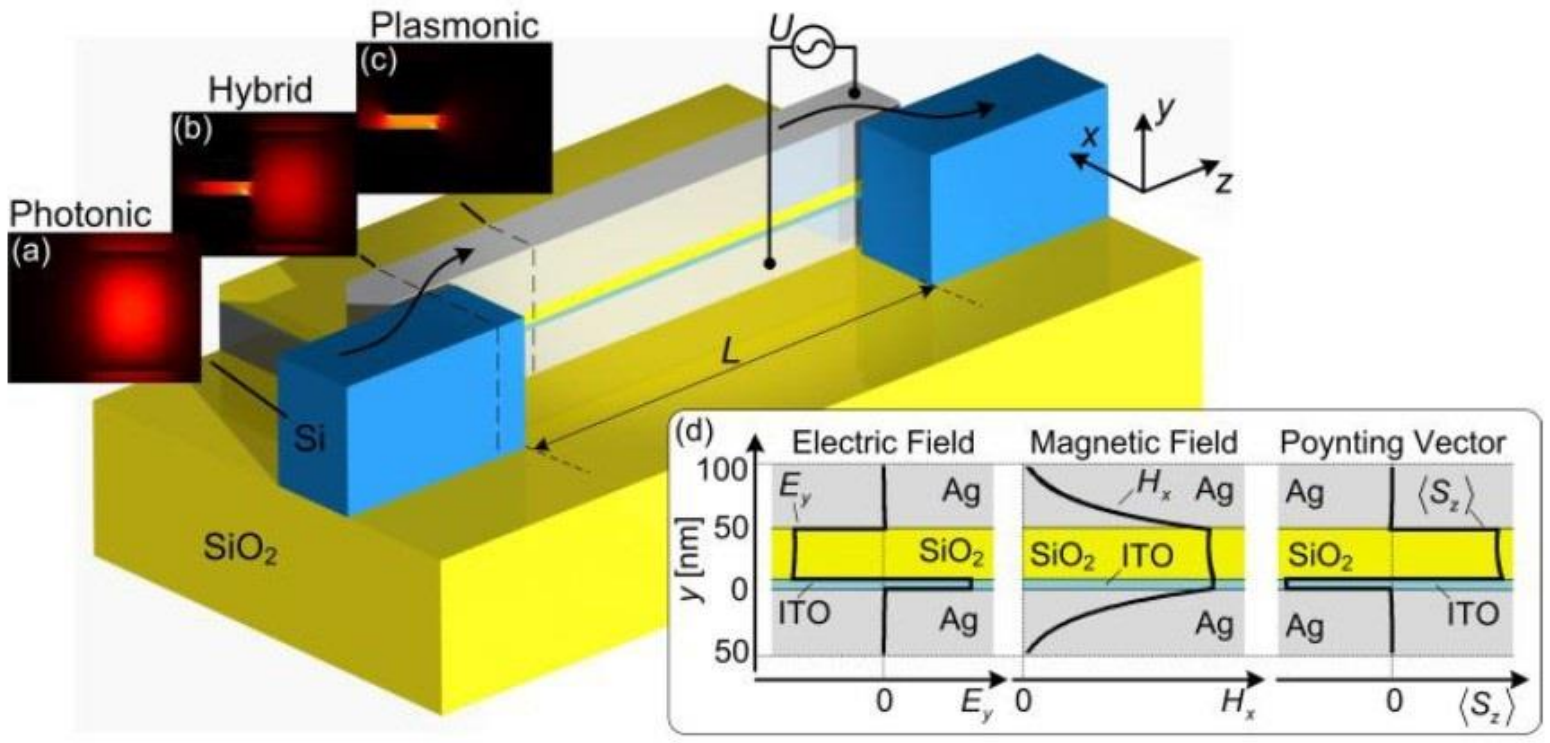

**b**

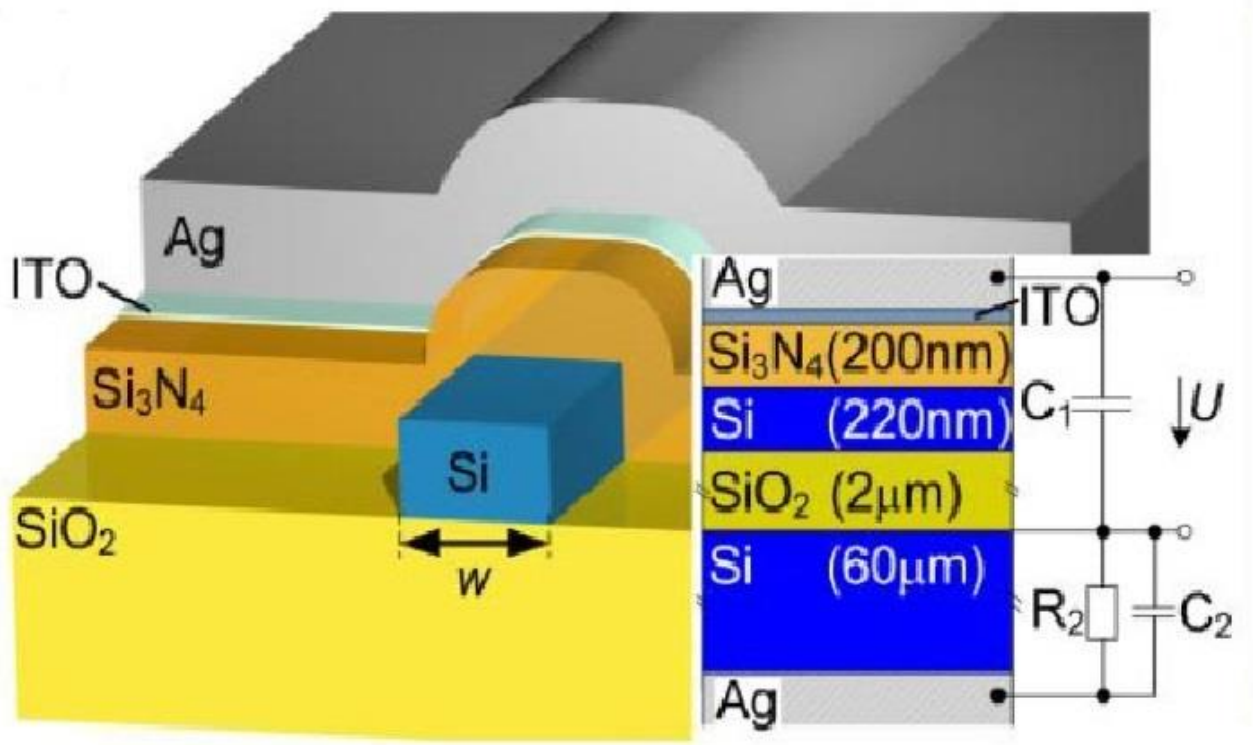

# Figure 9 of 14

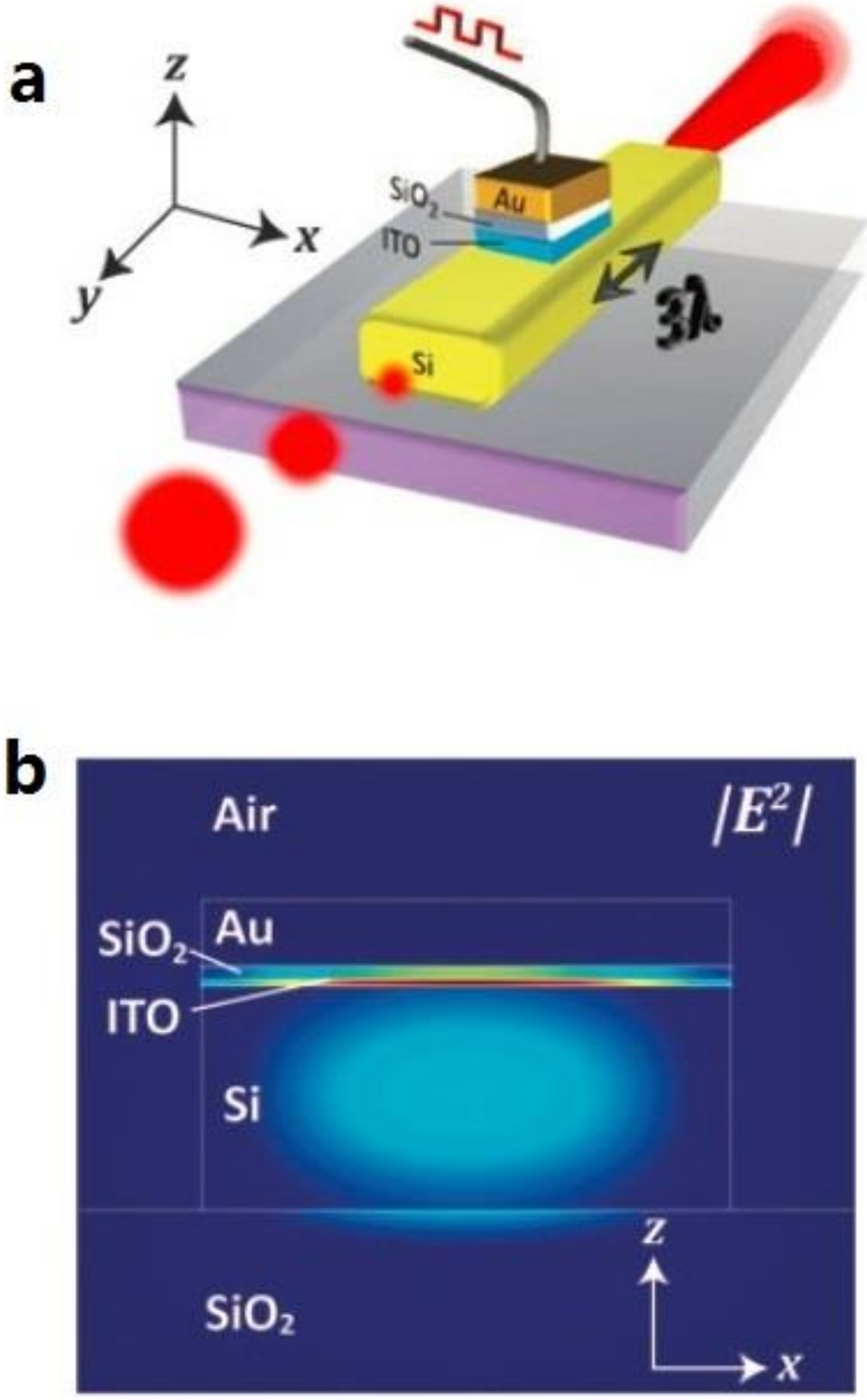

# Figure 10 of 14

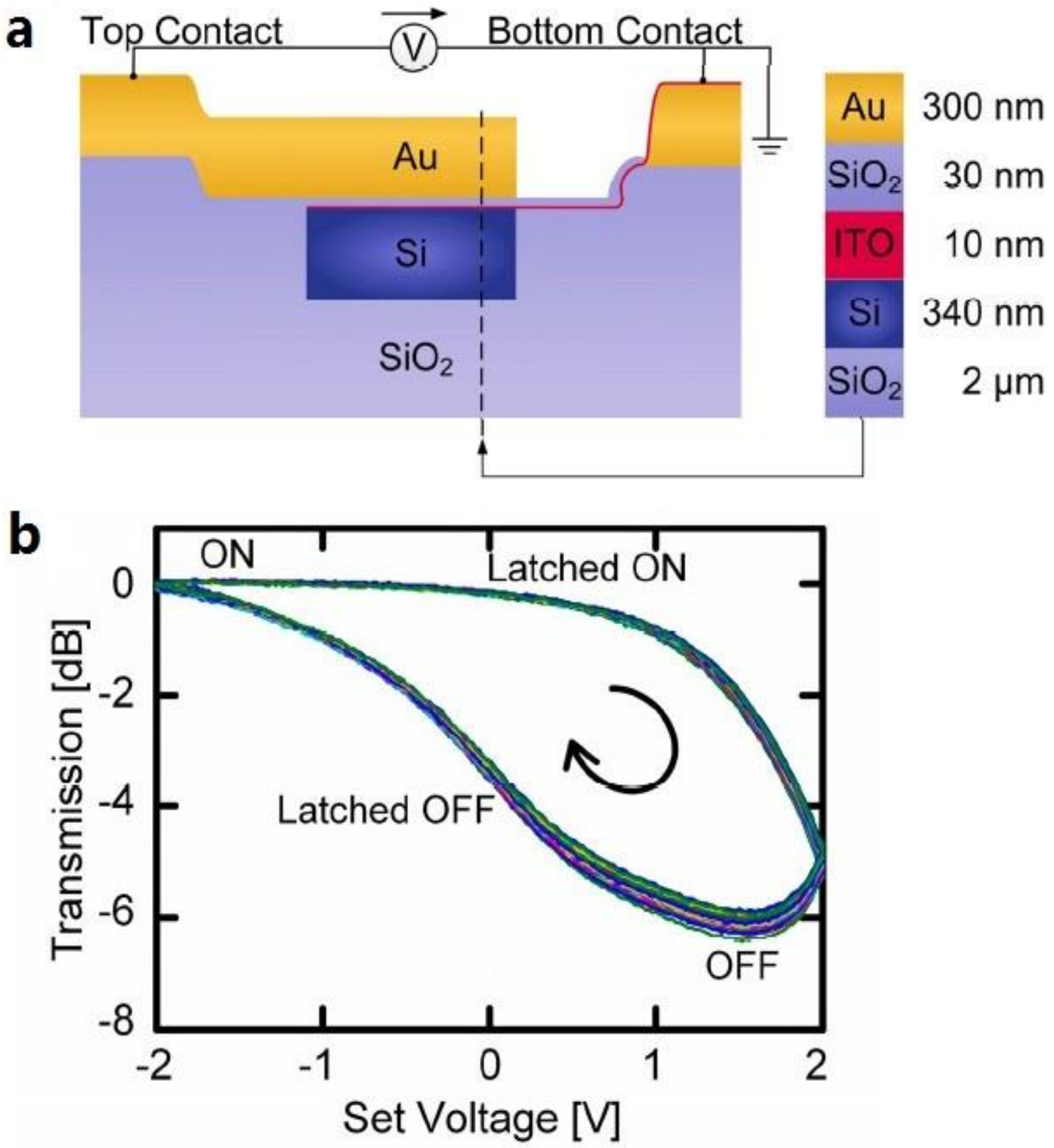

# Figure 11 of 14

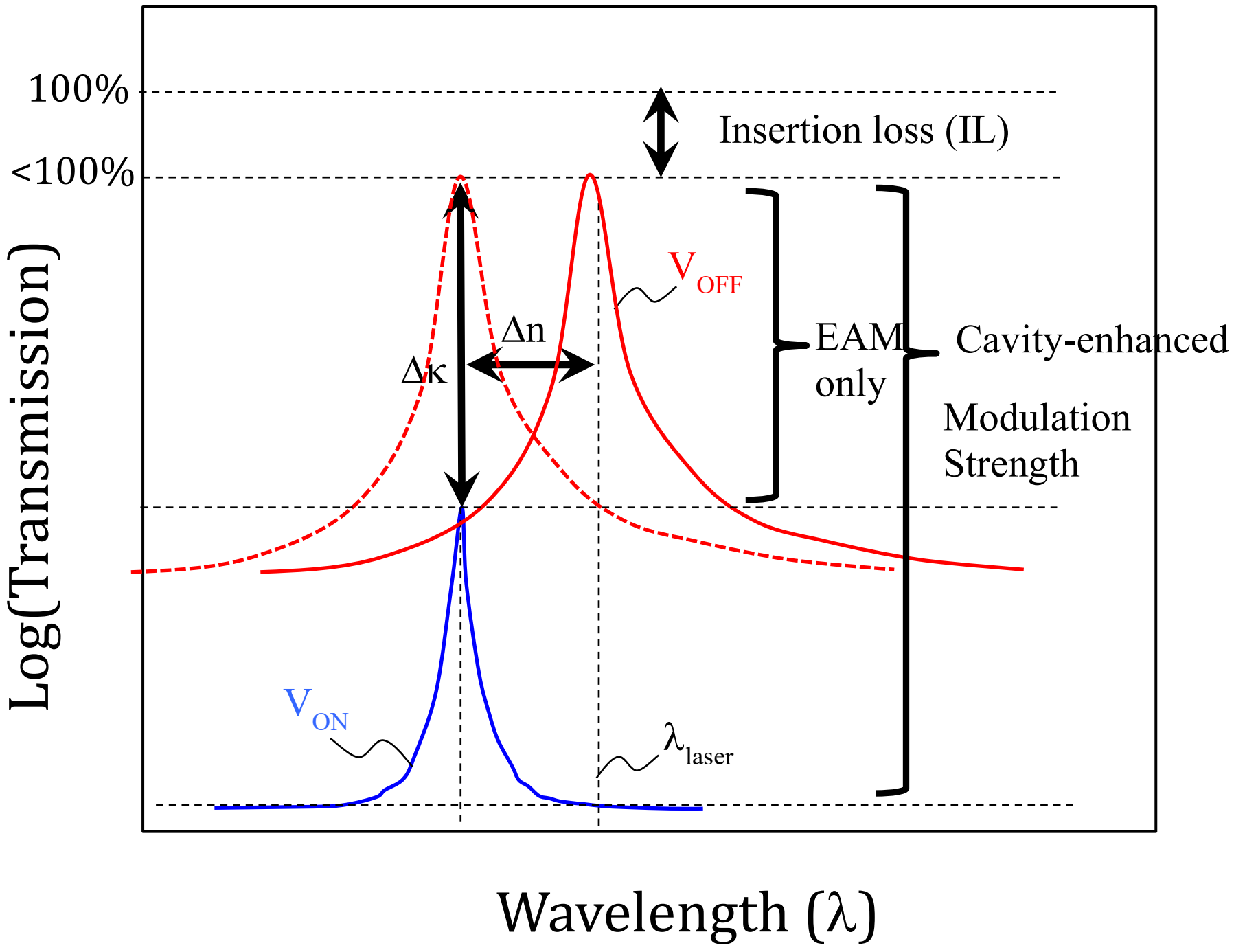

# Figure 12 of 14

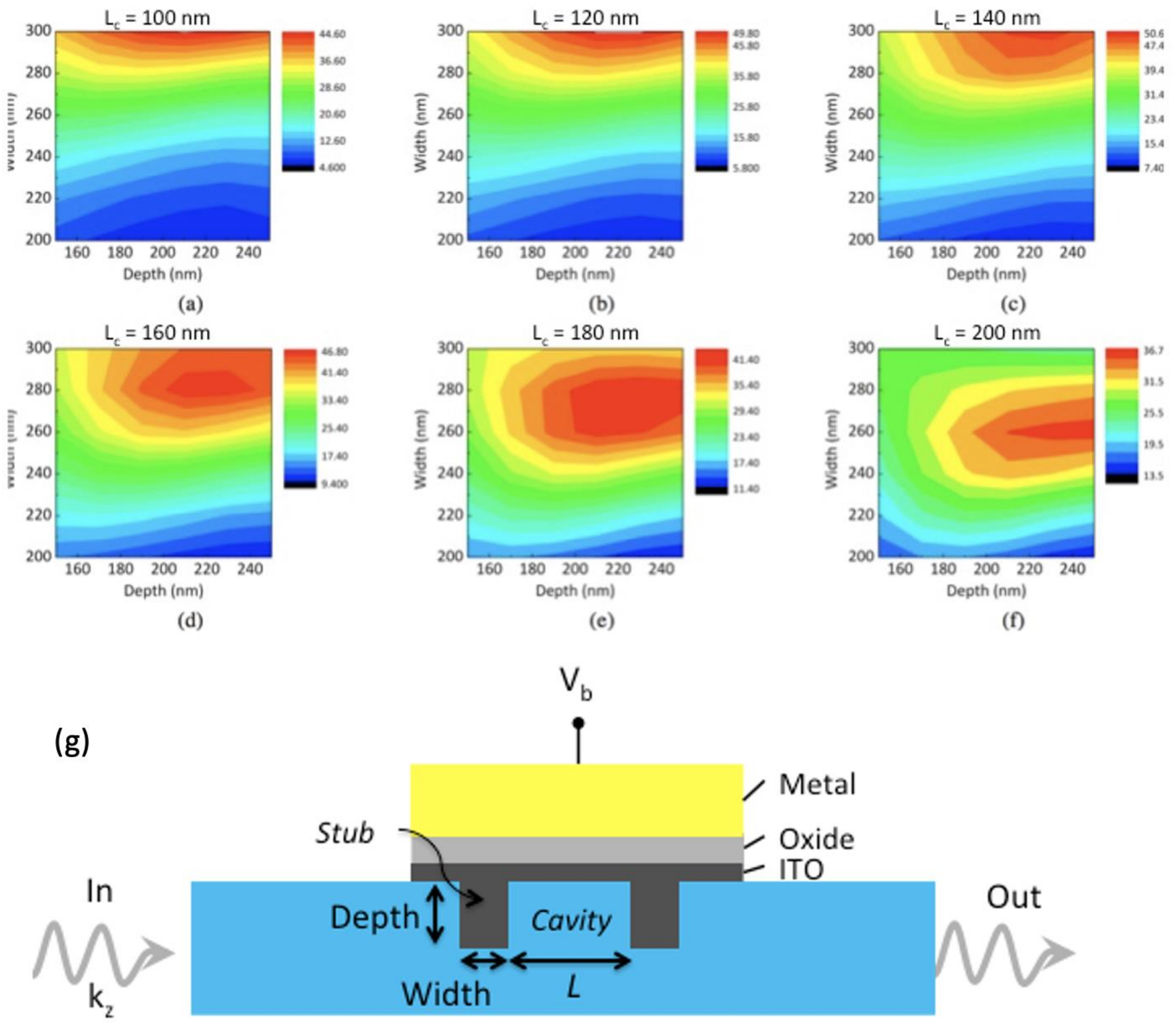

# Figure 13 of 14

**Table 2. Performance comparison for a non-cavity ITO-based EOM (row 1) and a Fabry-Perot based cavity.** The operation wavelength in the simulation is 1310 nm and the length of the device is 1 μm with the Si waveguide height being 250 nm. As a comparison, the device performance without a cavity are about 1/10 of those which feature the cavity. Results show an about 10x improvement in the FOM defined as extinction-ratio (ER) to the length normalized insertion-loss (IL) ratio.

| Cavity_L (nm) | Stub_W (nm) | Stub_d (nm) | IL (dB) | ER (dB) | ER/IL-um |
|---|---|---|---|---|---|
| / | / | / | 0.48 | 2.20 | 4.57 |
| 100 | 300 | 210 | 0.26 | 11.67 | 44.94 |
| 120 | 300 | 210 | 0.24 | 12.14 | 50.08 |
| 120 | 300 | 230 | 0.24 | 11.93 | 50.16 |
| 140 | 300 | 230 | 0.24 | 12.05 | 50.51 |
| 160 | 280 | 210 | 0.24 | 11.27 | 46.73 |
| 180 | 280 | 230 | 0.25 | 10.42 | 41.72 |
| 200 | 260 | 250 | 0.25 | 9.01 | 36.60 |

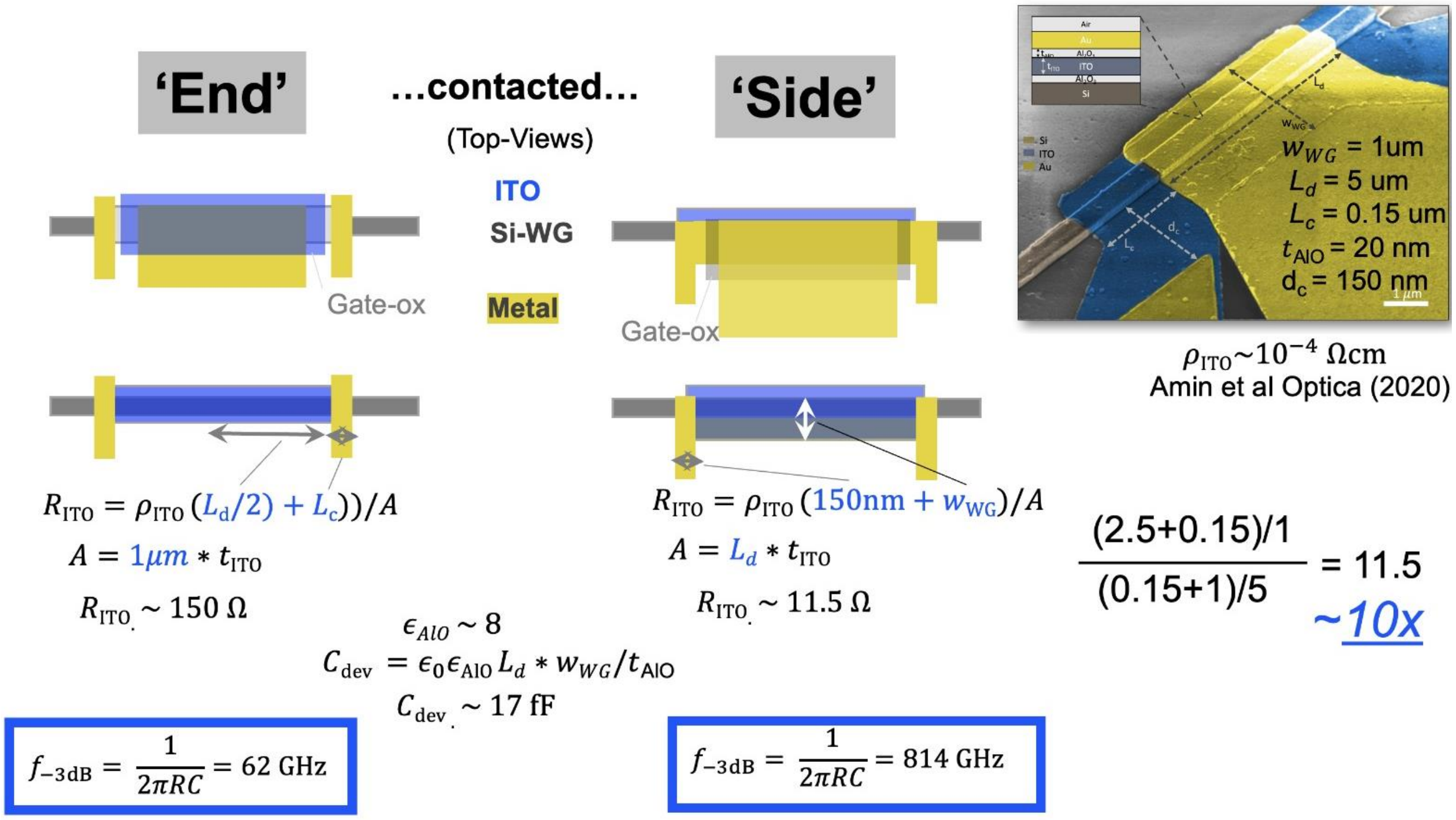

# Figure 14 of 14

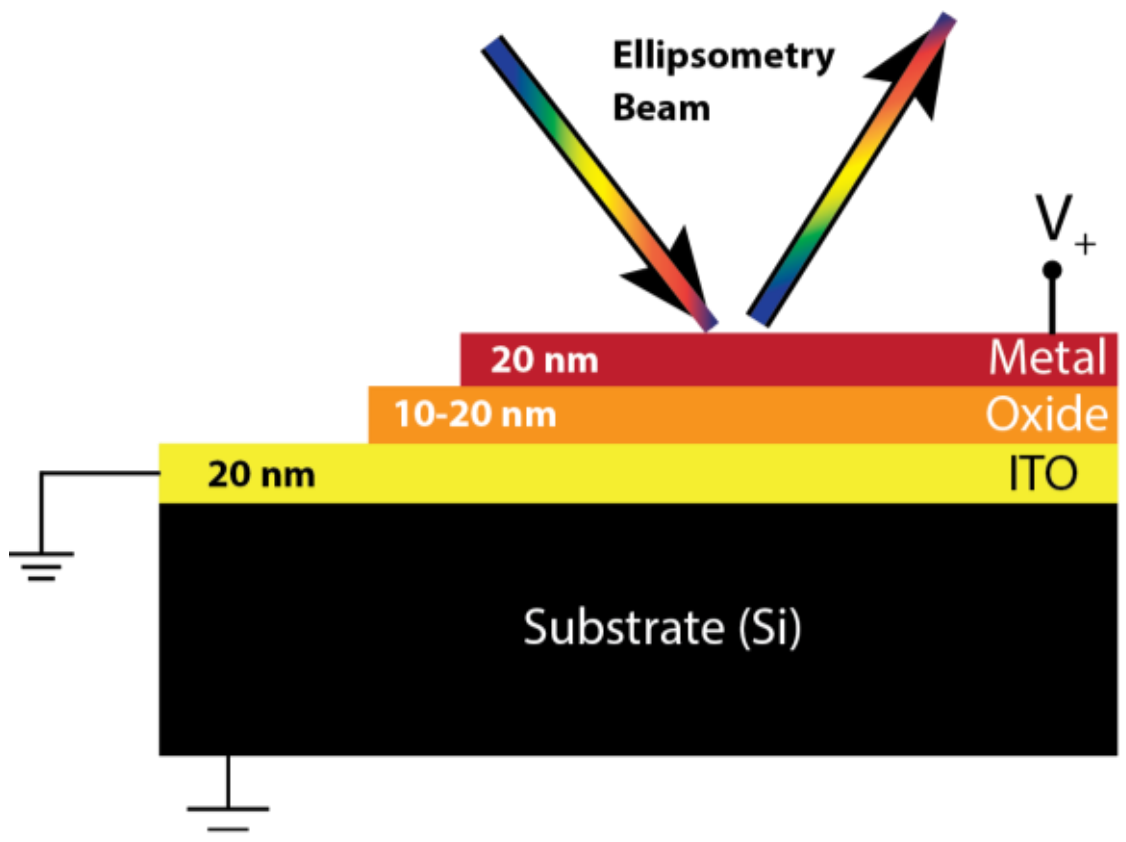

TOC Figure

**TOC Caption:**

The unique electro-optic properties of Indium Tin Oxide (ITO) such as unity refractive index modulation and epsilon-near-zero (ENZ) behavior allow for new functionality of active photonic devices on-chip. However processing conditions and physical non-concurrencies in recent work have let to question in the field. Here we review recent advanced of ITO processing and devices specifically for electro-optic modulation; we highlight promising results and point to open research directions.